\documentclass[twocolumn]{aastex62}

\usepackage{natbib}

\begin{document}

\title{SN 2017ein and the Possible First Identification of a Type Ic Supernova Progenitor}

\author{Schuyler D.~Van Dyk}
\affil{Caltech/IPAC, Mailcode 100-22, Pasadena, CA 91125, USA 0000-0001-9038-9950}

\author{WeiKang Zheng}
\affil{Department of Astronomy, University of California, Berkeley, CA 94720-3411, USA}

\author{Thomas G.~Brink}
\affil{Department of Astronomy, University of California, Berkeley, CA 94720-3411, USA}

\author{Alexei V.~Filippenko}
\affil{Department of Astronomy, University of California, Berkeley, CA 94720-3411, USA 0000-0003-3460-0103}
\affil{Miller Senior Fellow, Miller Institute for Basic Research in Science, University of California, Berkeley, CA 94720, USA}

\author{Dan Milisavljevic}
\affil{Department of Physics and Astronomy, Purdue University, 525 Northwestern Avenue, West Lafayette, IN 47907, USA}

\author{Jennifer E.~Andrews}
\affil{Steward Observatory, University of Arizona, 933 N.~Cherry Avenue, Tucson, AZ 85721, USA}

\author{Nathan Smith}
\affil{Steward Observatory, University of Arizona, 933 N.~Cherry Avenue, Tucson, AZ 85721, USA}

\author{Michele Cignoni}
\affil{Dipartimento di Fisica 'Enrico Fermi', Universit\'a di Pisa, largo Pontecorvo 3, I-56127 Pisa, Italy 0000-0001-6291-6813}
\affil{INFN, Largo B. Pontecorvo 3, I-56127 Pisa, Italy}

\author{Ori D.~Fox}
\affil{Space Telescope Science Institute, 3700 San Martin Drive, Baltimore, MD 21218, USA 0000-0002-4924-444X}

\author{Patrick L.~Kelly}
\affil{Department of Astronomy, University of California, Berkeley, CA 94720-3411, USA}
\affil{College of Science \& Engineering, Minnesota Institute for Astrophysics, University of Minnesota, 115 Union St.~SE, Minneapolis, MN 55455 USA 0000-0003-3142-997X}

\author{Angela Adamo}
\affil{Department of Astronomy, Oskar Klein Centre, Stockholm University, AlbaNova University Centre, SE-106 91 Stockholm, Sweden 0000-0002-8192-8091}

\author{Sameen Yunus}
\affil{Department of Astronomy, University of California, Berkeley, CA 94720-3411, USA}

\author{Keto Zhang}
\affil{Department of Astronomy, University of California, Berkeley, CA 94720-3411, USA}

\author{Sahana Kumar}
\affil{Department of Astronomy, University of California, Berkeley, CA 94720-3411, USA}
\affil{Department of Physics, Florida State University, 77 Chieftain Way, Tallahassee, Florida 32306, USA}

\begin{abstract}
We have identified a progenitor candidate in archival {\sl Hubble Space Telescope\/} ({\sl HST}) images for the 
Type Ic SN 2017ein in NGC 3938, pinpointing the candidate's location via {\sl HST\/} Target-of-Opportunity 
imaging of the SN itself. This would be the first identification of a stellar-like object as a progenitor candidate
for any Type Ic supernova to date. 
We also present observations of SN 2017ein during the first ${\sim}49$ days since explosion. 
We find that SN 2017ein most resembles the well-studied Type Ic SN 2007gr. We infer that SN 2017ein experienced a 
total visual extinction of $A_V \approx 1.0$--1.9 mag, predominantly because of dust within the host galaxy. Although 
the distance is not well known, if this object is the progenitor, it was likely of high initial mass, 
${\sim}47$--48 $M_{\odot}$ if a single star, or ${\sim}60$--80 $M_{\odot}$ if in a binary system. However, we also 
find that the progenitor candidate could be a very blue and young compact cluster, further implying a
very massive ($>65\ M_{\odot}$) progenitor. Furthermore, the actual progenitor might not be associated with the 
candidate at all and could be far less massive. From the immediate stellar environment, we find possible 
evidence for three different populations; if the SN progenitor was a member of the youngest population, this would be consistent 
with an initial mass of ${\sim}57\ M_{\odot}$. After it has faded, the SN should be reobserved at high spatial 
resolution and sensitivity, to determine whether the candidate is indeed the progenitor.
\end{abstract}


\section{Introduction}\label{intro}

Supernovae (SNe) are among the most powerful explosions in the Universe and highly influential within their host galaxies 
throughout cosmic time. Understanding their origins, as the catastrophic endpoints of stellar evolution, is therefore an important 
line of inquiry in modern astrophysics. Stars more massive than $\sim 8$--10 $M_{\odot}$ perish as core-collapse SNe 
(CCSNe). The most common CCSNe in the local Universe are the Type II-Plateau (SNe II-P), and 
we now have solid evidence that these arise from stars in the red supergiant phase \citep[e.g.,][]{Smartt+2009}.
The situation is less clear for SNe whose progenitor has had its H-rich envelope partially or entirely stripped away before
explosion, the so-called stripped-envelope SNe \citep[SESNe; e.g.,][]{Filippenko1997}.  We now have accumulating evidence 
that the progenitors of Type IIb SNe (SNe IIb), stripped but still retaining $\lesssim 0.1\ M_{\odot}$ of hydrogen, appear to be 
yellow to somewhat blue supergiants
\citep[e.g.,][]{Maund+2011,VanDyk+2011,VanDyk+2014,Folatelli+2015,Bersten+2018}, likely in interacting binary systems of 
moderate mass \citep[$\sim 12$--15 $M_{\odot}$; e.g.,][]{Podsiadlowski+1993,Stancliffe+2009,Benvenuto+2013}.

\bibpunct[;]{(}{)}{;}{a}{}{;}

For CCSNe which are hydrogen-free, there is only one identified progenitor of a SN Ib, for iPTF13bvn \citep{Cao+2013}, although
the actual nature of the star is still to be determined \citep{Folatelli+2016,Eldridge+2016}.
However, for the H-free SNe with little or no He as well, the SNe Ic, the progenitors to date have been elusive.
Here we are considering normal SNe Ic; more extreme broad-lined examples of SNe Ic have been found associated with
long-duration gamma-ray bursts \citep[e.g.,][]{Hjorth+2012}.
SN progenitor stars with binary companions that remain bound after explosion can evolve into many exotic configurations, 
including neutron star--neutron star mergers that are associated with short-duration gamma-ray bursts, kilonovae, and 
gravitational waves \citep[e.g.,][]{Abbott+2017}.

To remove most or all of the He from the progenitor, some mechanism for extensive mass loss is required.
\citet{Hachinger+2012} showed that it is indeed hard to hide H and He in such SNe, so the lack of such spectral features 
implies the presence of efficient mass loss.
One way is through a strong stellar wind from a single, massive star; for example, \citet{Georgy+2009} concluded that SNe Ic 
would result from the explosion of a highly massive, but stripped, star in the WC or WO Wolf-Rayet (WR) phase. 
\citet{Dessart+2012} suggested that the lack of observed He~{\sc i} lines in SN Ic spectra could arise from a high-mass,
possibly single, progenitor.
The other mass-loss mechanism is via envelope stripping in a mass-transfer binary system.
\citet{Nomoto+1990} modeled the SN Ic 1987M progenitor as a low-mass ($\sim 3$--3.5 $M_{\odot}$) He star, with initially
12--15 $M_{\odot}$, in a close interacting binary system.
A low-mass binary model was also invoked to explain the SN Ic 1994I \citep{Nomoto+1994}.
\citet{Yoon+2010} found a bimodality in their model SN Ic progenitors at solar metallicity, with lower-luminosity SNe (e.g., SN 
1994I) from lower-mass ($M_{\rm ZAMS} \approx 12$--13 $M_{\odot}$) binaries and a majority of SNe Ic from systems with 
higher-mass ($M_{\rm ZAMS} \gtrsim 33\ M_{\odot}$) primaries with WR wind mass loss. A similar progenitor system
dichotomy had been found earlier by \citet{Wellstein+1999} and \citet{Pols+2002}.
Many of the progenitors of SESNe must be lower-mass binaries, since the fraction of such SNe is locally too high to have
single-star progenitors (\citealt{Smith+2011}; see also \citealt{Graur+2017}, e.g., their Figure 10).
See also \citet{Zapartas+2017} for a discussion of pathways to SESNe.

The lack of detected progenitors for SNe Ic is not for a lack of attempting to locate them. 
Over the last two decades, a valiant effort has been expended by a number of investigators toward detecting SN Ic progenitors.
Their identification has been particularly thwarted by their proximity to luminous star clusters
(e.g., SNe 2004gt and 2013dk, both in the Antennae; \citealt{GalYam+2005,Maund+2005,EliasRosa+2013})
or high extinction (\citealt{Eldridge+2013}, who considered a number of SNe Ic), or both effects.
Other influences on our ability to detect SN Ic progenitors include the available archival pre-SN imaging 
\citep[e.g.,][]{Eldridge+2013}; moreover, massive WR stars, although bolometrically highly luminous, become optically less
luminous toward the end of their lives \citep{Yoon+2012}.
Additionally, archival images are often not obtained in filters sensitive to the strong, broad emission lines of WR stars, particularly 
He~{\sc ii} $\lambda$4686, at which WR stars are brighter than their continua by up to 3 mag 
\citep[][ see also Figure 7 of \citealt{Shara+2013}]{Massey+1998}.

The existing observational limits on SN Ic progenitors provide some informative constraints. Based on {\sl HST\/} data,
\citet{GalYam+2005} placed limits of $M_V>-5.5$ and $M_B>-6.5$ mag on the SN 2004gt progenitor and 
eliminated analogs of more than half of the known Galactic WR stars as possible progenitors.
Similarly, \citet{Maund+2005}, from the same dataset, limited the SN 2004gt progenitor to a low-luminosity, high-temperature star, 
such as a single high initial mass ($>40\ M_{\odot}$), carbon-rich WC star.
\citet{EliasRosa+2013} set a limit of $M_{\rm F555W} \gtrsim -5.7$ mag on the SN 2013dk progenitor luminosity and could not
rule out WR stars. The observational limits in the case of either SN did not provide constraints on lower-mass binary systems.
However, more recently, \citet{Johnson+2017} placed astoundingly low 1$\sigma$ upper limits of 
$M_U > -3.8$, $M_B > -3.1$, $M_V > -3.8$, and $M_R > -4.0$ mag on the presence of the SN Ic 2012fh progenitor in
deep ground-based images of the host galaxy, NGC 3344, ruling out essentially all single-star models.

\begin{figure}
\centering
\includegraphics[width=0.45\textwidth]{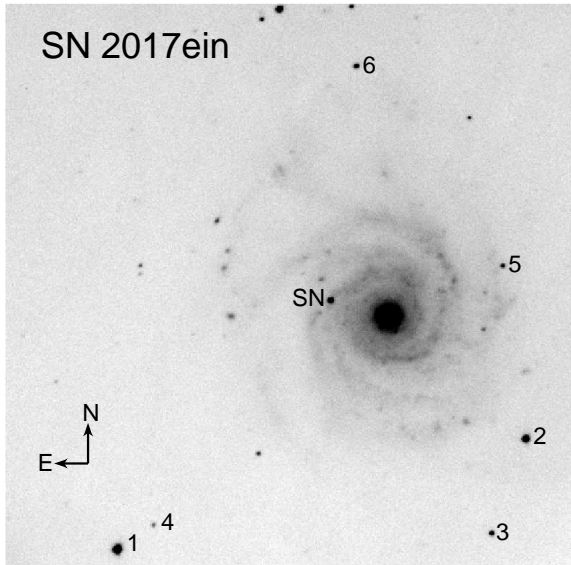}
\caption{KAIT unfiltered image, obtained on 2017 June 26.207, of SN 2017ein and the surrounding $6{\farcm}4 \times 6{\farcm}4$
field. The SN position is indicated, along with several stars that were used as local calibrators (see Table~\ref{tabseq}) for the 
KAIT and Nickel SN photometry.\label{figchart}}
\end{figure}

\begin{deluxetable}{ccccc}
\tablecolumns{5}
 \tablewidth{0pt}
 \tablecaption{Photometric Sequence for SN~2017ein\tablenotemark{a}\label{tabseq}}
 \tablehead{\colhead{Star} & \colhead{$B$} & \colhead{$V$} & \colhead{$R$} &  \colhead{$I$} \\
 \colhead{} & \colhead{(mag)} & \colhead{(mag)} & \colhead{(mag)} & \colhead{(mag)}}
 \startdata
1 & 14.383(034) & 13.842(014)	& 13.519(016) & 13.151(018) \\
2 & 15.844(034) & 15.065(015) & 14.613(017) & 14.195(019) \\
3 & 17.721(034) & 16.936(014)	& 16.481(016) & 16.044(018) \\
4 & 18.403(035) & 17.961(017)	& 17.691(019) & 17.385(021) \\
5 & 18.207(035) & 17.477(016) & 17.052(018) & 16.608(020) \\
6 & 18.093(033) & 17.003(013) & 16.380(016) & 15.818(018) \\
\enddata
\tablenotetext{a}{Uncertainties are provided in parentheses in thousandths of a magnitude.}
\end{deluxetable}

In this paper we describe what could be the first-ever identification and characterization of the progenitor of a SN Ic,
the nearby SN 2017ein in NGC 3938.
We first present early-time data on the SN itself and compare its properties to those of other well-studied SNe Ic, in particular SN 2007gr.
We then show our detection of a candidate for the progenitor in archival {\sl Hubble Space Telescope\/} ({\sl HST}) images,
which we isolated with high-spatial-resolution {\sl HST\/} images of the SN;
we initially reported this identification in \citet{VanDyk+2017}.
Finally, we attempt to constrain the nature of
the detected progenitor and compare these properties to predictions from the recent models of SESN progenitors.

SN 2017ein was discovered optically by \citet{Arbour2017} on May 25.99 (UT dates are used throughout this paper) at 
$\sim17.6$\,mag, and confirmed by {\sl Gaia\/} on May 29.71 at $G \approx 16.9$\,mag and given the designation 
Gaia17bjw\footnote{https://wis-tns.weizmann.ac.il/object/2017ein.}.
The SN was classified from a spectrum obtained on May 26.6 as Type Ic, although initially as broad-lined within one week of 
maximum light \citep{Xiang+2017}.
\citet{Im+2017}, through their regular monitoring of the host galaxy, were able to determine that the discovery by Arbour must
have been made shortly after explosion, since the SN was detectable in their images from May 25, but not from May 24.
\citeauthor{Im+2017}~found that the SN was continuing to rise in brightness in early June; thus, the initial classification
spectrum was most likely not obtained near maximum light.
We note that NGC 3938 was also host to the SN II-L 1961U \citep{Bertola1963}, the SN Ic 1964L \citep{Bertola+1965}, and 
the SN II-P 2005ay \citep{Tsvetkov+2006,GalYam+2008}.

\section{Observations}

\subsection{Ground-Based Imaging}

\begin{deluxetable*}{ccccccc}
\tablecolumns{7}
 \tablewidth{0pt}
 \tablecaption{KAIT and Nickel Photometry of SN~2017ein\tablenotemark{a}\label{tabphot}}
 \tablehead{\colhead{MJD} & \colhead{$B$} & \colhead{$V$} & \colhead{$R$} & \colhead{Unfiltered} & \colhead{$I$} & 
 \colhead{Source} \\
 \colhead{} & \colhead{(mag)}  & \colhead{(mag)}  & \colhead{(mag)}  & \colhead{(mag)}  & \colhead{(mag)}  & \colhead{}}
 \startdata
57902.27 & 17.18(16) & 16.49(19) & 16.19(20) & \nodata    & 15.92(22) & KAIT \\ 
57903.23 & 16.81(18) & 16.23(17) & 15.95(20) & 15.89(17) & 15.68(19) & KAIT \\
57903.23 & 16.72(01) & 16.23(06) & 15.89(07) & \nodata    & 15.63(07) & Nickel \\
57905.24 & 16.42(17) & 15.81(13) & 15.44(12) & 15.45(12) & 15.16(14) & KAIT \\ 
57906.22 & 16.18(16) & 15.63(13) & 15.33(14) & \nodata    & 15.01(15) & KAIT \\
57907.26 & 16.03(19) & 15.56(20) & 15.29(20) & 15.18(17) & 14.91(22) & KAIT \\
57908.25 & 16.04(18) & 15.39(17) & 15.21(20) & 15.11(20) & 14.77(20) & KAIT \\
57909.22 & 15.95(17) & 15.36(17) & 15.05(18) & \nodata    & 14.73(18) & KAIT \\
57910.20 & 15.84(08) & 15.27(07) & 14.96(12) & \nodata    & 14.63(10) & Nickel \\
57910.23 & 16.01(20) & 15.26(17) & 14.98(18) & 14.93(21) & 14.59(19) & KAIT \\
57917.21 & 16.51(27) & 15.32(21) & 14.87(21) & \nodata    & 14.52(22) & KAIT \\
57918.21 & 16.60(18) & 15.36(19) & 14.89(20) & 15.02(23) & 14.50(21) & KAIT \\
57919.21 & 16.62(18) & 15.41(15) & 14.92(16) & 15.01(17) & 14.50(18) & KAIT \\
57920.21 & 16.82(16) & 15.49(16) & 14.96(17) & 15.06(18) & 14.50(20) & KAIT \\
57922.21 & 16.98(14) & 15.64(14) & 15.08(16) & 15.17(11) & 14.57(17) & KAIT \\
57923.21 & 17.17(19) & 15.68(18) & 15.15(20) & 15.26(18) & 14.62(22) & KAIT \\
57924.21 & 17.20(16) & 15.79(14) & 15.21(16) & 15.31(22) & 14.68(16) & KAIT \\
57924.23 & 17.10(08) & 15.79(12) & 15.21(11) & \nodata    & 14.70(14) & Nickel \\
57925.21 & 17.24(14) & 15.90(13) & 15.27(15) & 15.34(11) & 14.73(15) & KAIT \\
57925.21 & 17.20(09) & 15.90(11) & 15.28(11) & \nodata     & 14.76(12) & Nickel \\
57926.23 & 17.40(17) & 15.96(17) & 15.37(18) & \nodata    & 14.80(19) & KAIT \\
57927.22 & 17.49(17) & 16.08(17) & 15.46(20) & 15.50(16) & 14.84(19) & KAIT \\
57928.21 & 17.47(08) & 16.13(09) & 15.48(10) & \nodata    & 14.90(11) & Nickel \\
57928.22 & 17.56(17) & 16.16(18) & 15.50(18) & 15.55(18) & 14.88(20) & KAIT \\
57929.19 & 17.62(24) & 16.13(18) & 15.56(17) & 15.58(17) & 14.96(19) & KAIT \\
57930.21 & 17.76(17) & 16.26(15) & 15.65(14) & 15.67(15) & 14.97(14) & KAIT \\
57931.23 & 17.78(20) & 16.34(17) & 15.61(19) & 15.68(16) & 14.99(19) & KAIT \\
57932.21 & 17.82(25) & 16.41(21) & 15.74(22) & 15.87(31) & 15.12(22) & KAIT \\
57933.21 & 17.83(16) & 16.46(13) & 15.81(12) & 15.82(15) & 15.13(13) & KAIT \\
57933.21 & 17.87(04) & 16.51(02) & 15.77(05) & \nodata    & 15.12(06) & Nickel \\
57934.19 & \nodata    & \nodata     & \nodata    & 15.84(17) & \nodata  &   KAIT \\
57938.19 & 18.13(27) & 16.76(16) & 16.08(15) & 16.09(15) & 15.32(15) & KAIT \\
57939.20 & 18.13(19) & 16.79(14) & 16.15(15) & 16.13(16) & 15.39(15) & KAIT \\
57940.19 & 18.32(42) & 16.78(19) & 16.15(19) & 16.24(21) & 15.42(20) & KAIT \\
57941.20 & 18.13(19) & 16.91(14) & 16.23(15) & 16.14(16) & 15.47(15) & KAIT \\
57942.19 & 18.10(35) & 16.86(16) & 16.21(13) & 16.20(14) & 15.49(16) & KAIT \\
57943.19 & 18.26(31) & 16.87(17) & 16.31(16) & 16.26(17) & 15.46(21) & KAIT \\
57944.19 & 18.43(38) & 16.92(18) & 16.30(19) & 16.24(16) & 15.58(21) & KAIT \\
57945.19 & 17.98(33) & 16.83(19) & 16.26(14) & 16.26(18) & 15.47(18) & KAIT \\
57947.20 & 18.22(23) & 16.98(14) & 16.30(12) & 16.29(12) & 15.51(15) & KAIT \\
57947.20 & 18.15(08) & 17.07(09) & 16.33(09) & \nodata     & 15.55(09) & Nickel \\ 
\enddata
\tablenotetext{a}{Uncertainties are provided in parentheses in hundredths of a magnitude.}
\end{deluxetable*}

SN 2017ein was constantly monitored photometrically in the bands $BVR_CI_C$ with both the 0.76-m Katzman Automatic 
Imaging Telescope \citep[KAIT;][]{Filippenko+2001} and the 1-m Nickel telescope at Lick Observatory, as well as with unfiltered 
KAIT images, beginning within 3 days of discovery and 
interrupted only by nights with poor observing conditions. We followed the SN until it was no longer accessible from either KAIT or
the Nickel telescope. We show a KAIT image of the field  in Figure~\ref{figchart}, with the SN indicated.
From our KAIT data we measured an absolute position for the SN of 11h 52m 53.26s $+44\arcdeg\ 07\arcmin 26{\farcs}2$
(J2000).
All images were reduced using a custom pipeline \citep{Ganesh+2010}.
Point-spread function (PSF) photometry was then obtained using DAOPHOT \citep{Stetson1987}
from the IDL Astronomy UserÕs Library\footnote{http://idlastro.gsfc.nasa.gov/.}.
We also tried to perform the photometry after using the image template-subtraction method, in order to estimate the host contribution 
to the light. We find that the differences between using subtraction and not using subtraction are very small ($<0.1$\,mag) for all  epochs, implying that the host contribution 
is minimal. We thus adopted the results without template subtraction.

Six stars in the KAIT and Nickel fields were chosen as local calibrators; they are indicated in Figure~\ref{figchart}. This field
was observed by Pan-STARRS and included in the release of the Mean Object 
Catalog\footnote{Available at https://archive.stsci.edu/panstarrs/.}. We obtained from that release the mean PSF photometry 
for the six stars in the PS1 bands and used the transformations to $BVR_CI_C$ provided by \citet{Tonry+2012}. We list the 
resulting magnitudes for the stars in Table~\ref{tabseq} and use these for calibration of the SN photometry.

Apparent magnitudes of the SN and the six calibrators were all measured in the KAIT4/Nickel2 natural system.
The final results were transformed to the standard system, using the local calibrators and color terms for KAIT4 
\citep[][ their Table 4]{Ganesh+2010} and updated Nickel color terms \citep{Shivvers+2017}.
We present the early-time KAIT and Nickel photometry of SN 2017ein in Table~\ref{tabphot}.

\begin{deluxetable*}{cccccc}
\tablecaption{Log of Spectroscopic Observations of SN 2017ein\label{tabspec}}
\tablecolumns{6}
\tablewidth{0pt}
\tablehead{
\colhead{UT date} & \colhead{MJD} & \colhead{Age\tablenotemark{a}} & \colhead{Instrument} & \colhead{Wavelength} & \colhead{Resolution} \\
\colhead{} & \colhead{} & \colhead{} & \colhead{} & \colhead{Range (\AA)} & \colhead{(\AA)}}
\startdata
2017 Jun 02.26 & 57906.76 & $-$6.3 & Kast & 3640--10,630 & 2.0 \\
2017 Jun 21.27 & 57925.77 & 12.7 & Kast & 3626--10,710 & 2.0 \\
2017 Jun 24.19 & 57928.69 & 15.6 & MMT &  5711--7022 & 0.5 \\
2017 Jun 27.26 & 57931.76 & 18.7 & Kast & 3630--10,712 & 2.0 \\
2017 Jul 01.24 & 57935.74 & 22.6 & Kast & 3636--10,710 & 2.0 \\
2017 Jul 17.21 & 57951.71 & 38.6 & Kast & 3612--10,700 & 2.0 \\
2017 Jul 26.20 & 57960.70 & 47.6 & Kast & 3622--10,670 & 2.0 \\
2017 Jul 30.20 & 57964.70 & 51.6 & Kast & 3620--10,704 & 2.0 \\
2017 Aug 01.19 & 57966.69 & 53.6 & Kast & 3620--10,680 & 2.0 \\
\enddata
\tablenotetext{a}{Day since estimated time of $V$ maximum, approximately 2,457,913.1 (June 8.6).}
\end{deluxetable*}

\subsection{Spectroscopy}

Over a two-month period beginning on 2017 June 2, eight optical spectra of SN 2017ein were obtained with the Kast 
Spectrograph mounted on the 3-m Shane telescope \citep{Miller+1993} at Lick Observatory.  They were taken 
at or near the parallactic angle \citep{Filippenko1982} to minimize slit losses caused by atmospheric dispersion. Data were 
reduced following standard techniques for CCD processing and spectrum extraction \citep{Silverman+2012} utilizing 
IRAF\footnote{IRAF is distributed by the National Optical Astronomy Observatory, which is operated by AURA, Inc., under a 
cooperative agreement with the NSF.} routines and custom Python and IDL 
codes\footnote{https://github.com/ishivvers/TheKastShiv.}. Low-order polynomial fits to comparison-lamp spectra 
were used to calibrate the wavelength scale, and small adjustments derived from night-sky lines in the target frames were 
applied. Observations of appropriate spectrophotometric standard stars were used to flux calibrate the spectra. 

We obtained $3 \times 1200$~s exposures with the Blue Channel spectrograph on the MMT on 2017 June 24 (JD 2,457,928.69). 
The data were taken with the 1200\,lines mm$^{-1}$ grating with a central wavelength of 6360\,\AA\ and a 1$\farcs$0 slit width. 
The seeing was 1$\farcs$2.   
Standard reductions were carried out using IRAF, and wavelength solutions were determined using internal He-Ne-Ar lamps. 
Flux calibration was achieved using spectrophotometric standards at a similar airmass taken throughout the night.
A log of all spectroscopic observations is provided in Table~\ref{tabspec}.

We note that we cross-checked the calibrations of our photometry and our spectra at four 
nearly contemporaneous epochs (MJD 57906.76, 57925.77, 57931.76, and 57935.74) by comparing observed colors with colors 
synthetically generated from the spectra  
using pysynphot\footnote{https://github.com/spacetelescope/pysynphot.}, and found that the two datasets differed by at most
$\sim 0.1$\,mag, within the uncertainties of the photometry; thus, we are confident that our calibrations are sufficient.

\subsection{{\sl HST\/} Imaging}

The SN 2017ein site is in publicly available archival images obtained with the 
{\sl HST\/} Wide Field Planetary Camera 2 (WFPC2) on 2006 October 20 (by program
GO-10877, PI A.~Filippenko, originally to observe SN 2005ay) in bands F555W (total exposure time 460 s) and F814W (700 s).
The image mosaics were obtained from the Hubble Legacy Archive, whereas the individual WFPC2 frames were obtained from 
the {\sl HST\/} Archive at MAST.

The SN site is just off the edge of the field of view (NIC3) of archival Near-Infrared Camera and Multi-Object Spectrograph 
(NICMOS) images obtained on 2003 April 3 in bands F160W, F187N, and F190N, and thus we do not consider these.

We also observed the transient itself with {\sl HST\/} on 2017 June 12 using the Wide-Field Camera 3 (WFC3) UVIS channel in 
subarray mode, as part of our Target of Opportunity (ToO) program (GO-14645, PI S.~Van Dyk),
in F438W, with 10\,s individual frame times and total exposure time 270\,s. 
The F438W band was chosen in this case to attempt to avoid saturation by the SN, at which we were successful.
All of these data had been initially processed via the default pipelines at STScI and were obtained from the {\sl HST\/} Archive.
The individual {\it flc\/} frames, which had been corrected by the pipeline for charge-transfer efficiency losses, were individually
combined into an image mosaic using AstroDrizzle within PyRAF.

\begin{figure}
\plotone{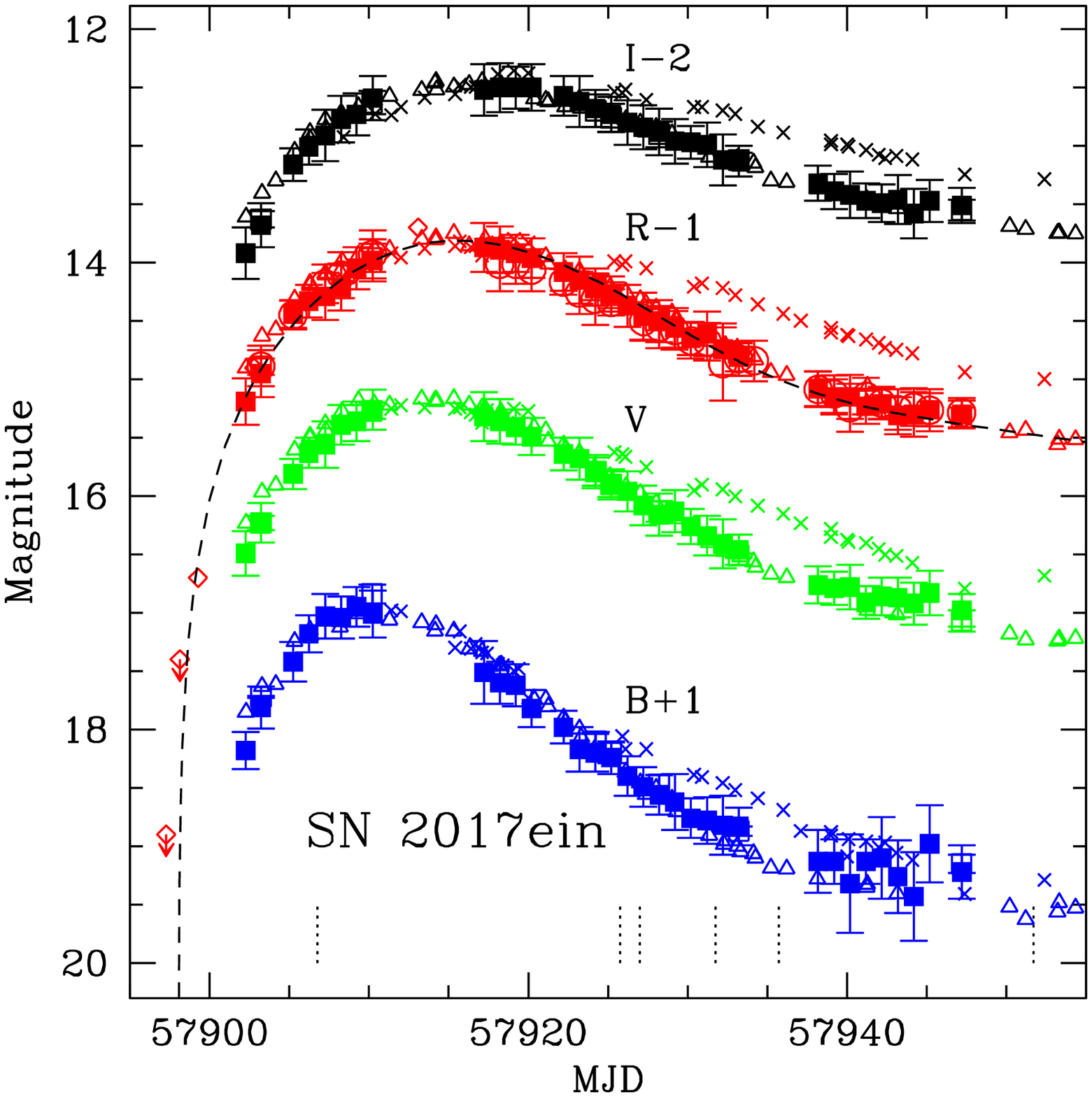}
\caption{Early-time optical KAIT (solid squares) and Nickel (open squares) $BVR_CI_C$ and KAIT unfiltered (open circles) light 
curves of SN 2017ein. Also shown are $R$ observations by \citet[][ open diamonds]{Im+2017}.
Additionally, for comparison we show the light curves for the SNe Ic 2007gr \citep[open triangles;][]{Hunter+2009}
and 2004aw \citep[crosses;][]{Taubenberger+2006}. 
The curves
for both SNe were shifted in time to match approximately the $V$ maximum for SN 2017ein. After correction for reddening
appropriate for these two SNe, their light curves were then reddened by the Galactic foreground for SN 2017ein and an 
additional host contribution of $A_V=1.05$ mag with $R_V=3.1$. The SN 2007gr light curves were further adjusted by a
difference in distance modulus of 1.32 mag, which is consistent with the difference between the distances to the SN 2007gr and
SN 2017ein hosts, to within the uncertainties; the SN 2004aw curves were adjusted by a distance modulus difference of 2.02
mag, which is a smaller difference than implied by the distance given in \citeauthor{Taubenberger+2006}
The dotted lines indicate the epochs of SN 2017ein spectroscopy.\label{figlc}}
\end{figure}

\begin{figure}
\plotone{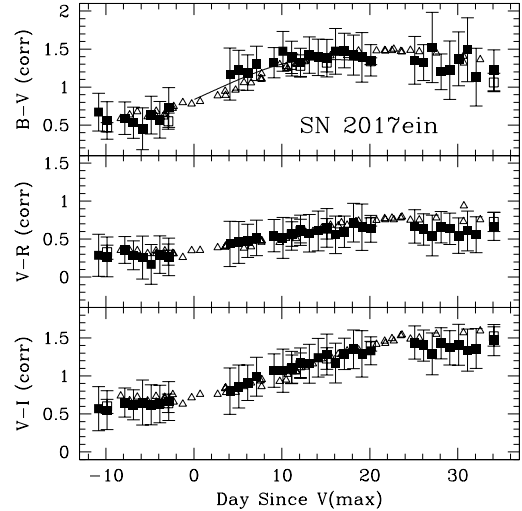}
\caption{Early-time $B-V$, $V-R$, and $V-I$ color curves of SN 2017ein (solid squares), after initial correction for reddening 
attributed to the Galactic foreground contribution \citep{Schlafly+2011}. For comparison we show the color curves of SN Ic 2007gr 
\citep[open triangles;][]{Hunter+2009}, corrected for Galactic foreground reddening and then reddened and shifted in time to 
match the curves of SN 2017ein. Additionally, we show the best fit of the SN Ic color template at $B-V$ from 
\citet[][ solid curve]{Stritzinger+2018}. See discussion in Section~\ref{extinction}.\label{figcolor}}
\end{figure}

\begin{figure}
\plotone{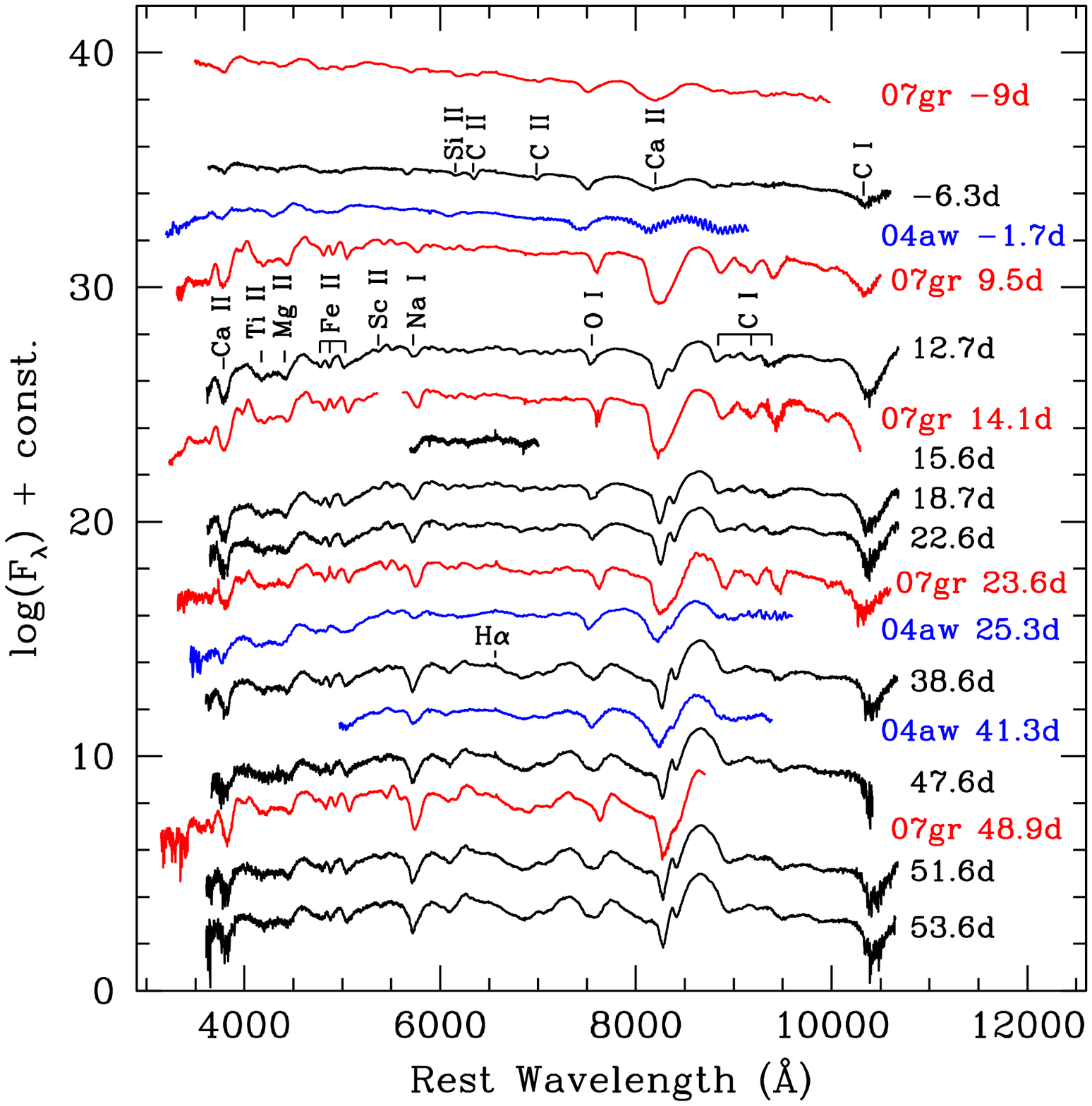}
\caption{Early-time optical spectra obtained with the Kast spectrograph on the 3-m Shane telescope at Lick 
Observatory (plus one narrow-range MMT spectrum on day 15.6). The ages are relative to $V$ maximum.
For comparison we show spectra of SN 2007gr \citep{Valenti+2008,Chen+2014} and
SN 2004aw \citep{Taubenberger+2006} at roughly similar ages.
Various absorption lines and features seen in the spectra are indicated.
The SN 2017ein spectra shown in this figure are available in the electronic journal and are also posted on 
WISEReP, https://wiserep.weizmann.ac.il/.\label{figspec}}
\end{figure}

\section{The Supernova: Early Results}

\subsection{Photometry}\label{secphot}

We show the combined KAIT and Nickel light curves in Figure~\ref{figlc}. Here we have also included the photometry at $R$ from
\citet{Im+2017}. We compared the light curves with those of a number of SNe Ic
and found that the best match was with the light curves from \citet{Hunter+2009} for the normal SN Ic 2007gr 
in NGC 1058. 
The light curves for SNe 2017ein and 2007gr show quite similar behavior.
We also compared the SN 2017ein light curves with those of SN 2004aw \citep{Taubenberger+2006}, which 
may agree at early times
but diverge post-peak, with SN 2004aw being generally more luminous at all bands.
(We have made the comparison with SN 2004aw here, since, as we show in Section~\ref{bolo}, SNe 2017ein and 2004aw could
be similar bolometrically.)
One can see that we missed observations of maximum light at $V$ [$V$(max)] for SN 2017ein, owing to poor observing 
conditions. Based on the comparison with SN 2007gr, we estimate, however, that $V$(max) occurred on 
JD $\approx$ 2,457,913.1 (i.e., June 8.6).

That the SN was not detected to $R>18.4$ mag in images obtained by \citet{Im+2017} on May 24.63 (JD 2,457,898.13), but then 
was seen at $R=17.7$ mag by these investigators on May 25.77 (JD 2,457,899.27), and later discovered by Arbour on May 
25.99, would imply that the explosion date was likely sometime between JD 2,457,898 and 2,457,899.
We analyzed the $R$-band light curve, including the \citet{Im+2017} points, by fitting a simple analytic model for H-free
SNe \citep[][ see Figure~\ref{figlc}]{Vacca+1997}, and found that the explosion could have been as early as JD 
2,457,898.1, or May 24.6, which we adopt.
The analytic model further implies that $R$-band maximum (again, missed by our observations) occurred on approximately JD 
2,457,916, or $\sim 3$ days after $V$-band maximum.

In Figure~\ref{figcolor} we show the early-time $B-V$, $V-R$, and $V-I$ color curves. We have initially corrected the observed 
color curves for Galactic foreground reddening \citep[][ via the NASA/IPAC Extragalactic Database, NED]{Schlafly+2011}.
Again, comparing with SN 2007gr \citep{Hunter+2009}, we can see that the color evolution of SN 
2017ein follows much the same behavior, to within the uncertainties. 
We also found that $V-R$ of SN 2017ein evolved at early times in a fashion similar to that of SN 2004fe 
\citep[not shown;][]{Drout+2011}.

\subsection{Spectroscopy}

We show the early-time spectra of SN 2017ein in Figure~\ref{figspec}. The first spectrum was obtained on June 2, about one
week after discovery (and explosion). From the light-curve comparison with SN 2007gr, we 
estimate the ages of the spectra relative to $V$(max). In the figure we also show a comparison with
spectra of SN 2007gr from \citet{Valenti+2008} and \citet{Chen+2014}, and of SN 2004aw from \citet{Taubenberger+2006},
at approximately the same ages.
The spectra of SN 2004aw and the \citeauthor{Valenti+2008}~spectra of SN 2007gr were obtained from 
WISeREP\footnote{https://wiserep.weizmann.ac.il/.} \citep{Yaron+2012}, and the \citeauthor{Chen+2014}~SN 2007gr 
spectra were obtained from the UC Berkeley Supernova Database\footnote{http://heracles.astro.berkeley.edu/sndb/.
\citep[SNDB;][]{Silverman+2012}.}
One can see that the spectra of SN 2017ein resemble quite closely those of SN 2007gr, as was also true for the light curves. 
SN 2017ein also somewhat resembles SN 2004aw, spectroscopically, although the latter may have had somewhat broader lines.
Many of the spectral features, which we have indicated in the figure, are the
same between SN 2017ein and SN 2007gr, and the line widths are comparatively narrow. 
SN 2017ein, like SN 2007gr, is clearly not a broad-lined SN Ic.

In the pre-maximum spectrum, the features at 6350\,\AA\ and at $\sim7000$\,\AA, attributed to 
C~{\sc ii} $\lambda$6580 and $\lambda$7235 in SN 2007gr \citep{Valenti+2008}, are evident in SN 2017ein, and the
Ca~{\sc ii} $\lambda\lambda$8542, 8662, 8498 near-infrared (NIR) feature and, 
interestingly, the $\sim10,400$\,\AA\ feature, attributed primarily to C~{\sc i}, both make an early appearance. 
In the later spectra, Ca~{\sc ii}, Ti~{\sc ii}, Mg~{\sc ii} $\lambda$4481, 
the well-resolved Fe~{\sc ii} $\lambda\lambda$4924, 5018, 5169, Sc~{\sc ii}, Na~{\sc i}, O~{\sc i} $\lambda$7774, and 
C~{\sc i} are also seen, as in SN 2007gr.
For both SN 2017ein and SN 2007gr at later times, weak interstellar H$\alpha$ emission appears.

\begin{figure}
\plotone{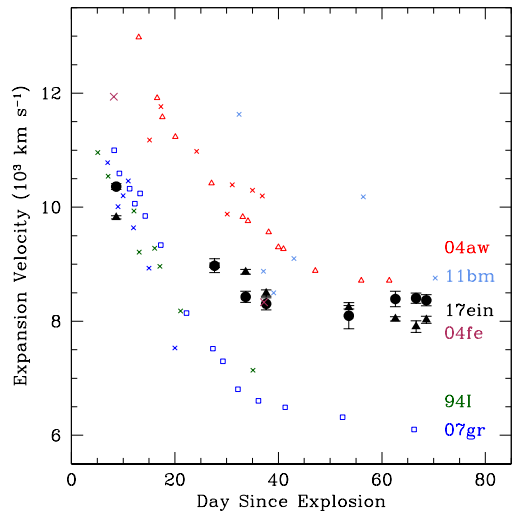}
\caption{Expansion velocity evolution of SN 2017ein, measured from the Fe~{\sc ii} $\lambda\lambda$4924, 5018, 5169 lines
(filled squares) and O~{\sc i} $\lambda$7774 (filled circles). 
For comparison we show the 
velocities of SN 2007gr from the Fe~{\sc ii} lines \citep[][ open squares]{Chen+2014}, SN 2004aw from the O~{\sc i} $\lambda$7774 line \citep[][ open triangles]{Taubenberger+2006}, and several SNe~Ic in the sample from \citet[][ crosses]{Liu+2016} from Fe~{\sc ii} lines.\label{figvel}}
\end{figure}

We measured (see Table~\ref{tabvel}) the photospheric expansion velocity of SN 2017ein 
from the Fe~{\sc ii} $\lambda\lambda$4924, 5018, 5169 and O~{\sc i} $\lambda$7774 absorption lines, by fitting a Gaussian to 
each of the lines with the routine {\it splot\/} in PyRAF. The measurements from the Fe~{\sc ii} lines were averaged.
We show these velocities for both sets of lines in Figure~\ref{figvel}, with a comparison to 
SN 2007gr \citep{Chen+2014}, SN 2004aw \citep{Taubenberger+2006}, 
and several SNe~Ic in the sample from \citet{Liu+2016}.
The expansion velocities of SN 2017ein are not initially (soon after explosion) as high at those of
SN 2007gr and SN 2004aw, and overall appear to be intermediate between 
these two other SNe.

Unlike SN 2007gr, though, as the Ca~{\sc ii} NIR triplet in SN 2017ein appeared to narrow in width with advancing age, a notch
of additional absorption appeared in the red wing of the triplet feature. 
It is unclear to what to attribute this feature; it could be the Ca~{\sc ii} $\lambda$8664 line becoming more apparent as the
triplet feature narrowed, or it could be the C~{\sc i} $\lambda$8727 line, which is normally blended with the Ca~{\sc ii} NIR triplet,
as the other C~{\sc i} absorption features also become stronger.

\begin{figure}
\plotone{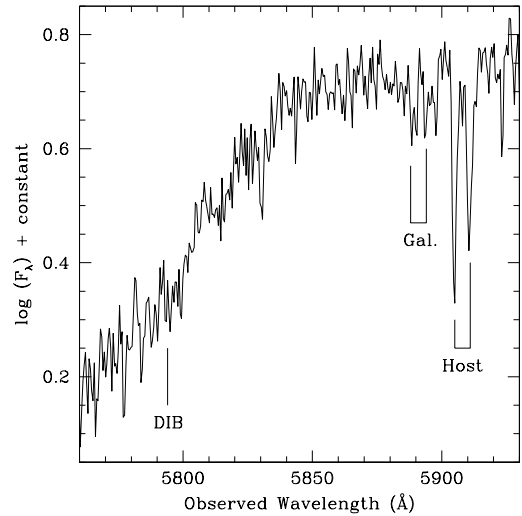}
\caption{A moderate-resolution spectrum of SN 2017ein obtained at the MMT on 2017 June 24. The interstellar
Na~{\sc i} D doublet at the expected wavelengths for both the Milky Way (``Gal.'') and NGC 3938 (``Host'') are indicated. Also
indicated is the expected location of the diffuse interstellar band (``DIB'') near rest wavelength 5780\,\AA\ for the host 
galaxy.\label{fighires}}
\end{figure}

Also notably exceptional for SN 2017ein is the presence of significant Na~{\sc i}~D absorption in the June 2 spectrum
(day $-6.3$), implying more interstellar extinction to SN 2017ein than to SN 2007gr.
This is particularly seen in Figure~\ref{fighires}, based on the MMT spectrum, in which the well-resolved Na~{\sc i} doublet is 
strong (at 5904.9 and 5910.6 \AA, respectively), attributable internally to the host galaxy. The doublet is far weaker in the Galactic
foreground, consistent with the assumed low reddening from this contribution.
We have also indicated in Figure~\ref{fighires} the presumed location from the host galaxy of the diffuse interstellar band (DIB) 
feature at $\lambda$5780,
the strength of which can be used to infer visual extinction $A_V$ \citep{Phillips+2013} and which can change over time in a
subset of SNe Ib/c \citep{Milisavljevic+2014}. 

\begin{deluxetable}{cccc}
\tablecaption{Photospheric Expansion Velocities for SN 2017ein\label{tabvel}}
\tablecolumns{4}
\tablewidth{0pt}
\tablehead{
\colhead{UT date} & \colhead{MJD} & \colhead{$v_{\rm exp}$ (Fe~{\sc ii})} & \colhead{$v_{\rm exp}$ (O~{\sc i})} \\
\colhead{} & \colhead{} & \colhead{(km s$^{-1}$} & \colhead{(km s$^{-1}$)}}
\startdata
2017 Jun 02.26 & 57906.76 & $9822 \pm 28$ & $10362 \pm 49$ \\
2017 Jun 21.27 & 57925.77 & $8383 \pm 37$ & $8973 \pm 122$ \\
2017 Jun 27.26 & 57931.76 & $8866 \pm 42$ & $8427 \pm 98$ \\
2017 Jul 01.24  & 57935.74 & $8494 \pm 58$ & $8306 \pm 105$ \\
2017 Jul 17.21  & 57951.71 & $8243 \pm 30$ & $8097 \pm 230$ \\
2017 Jul 26.20  & 57960.70 & $8042 \pm 37$ & $8391 \pm 137$ \\
2017 Jul 30.20  & 57964.70 & $7905 \pm 101$ & $8405 \pm 89$ \\
2017 Aug 01.19 & 57966.69 & $8027 \pm 61$ & $8368 \pm 97$ \\
\enddata
\end{deluxetable}

\subsection{Extinction to SN 2017ein\label{extinction}}

\begin{deluxetable}{ccc}
\tablecaption{Summary of Host Extinction Estimates for SN 2017ein\label{tabext}}
\tablecolumns{3}
\tablewidth{0pt}
\tablehead{
\colhead{Method} & \colhead{$A_V$} & \colhead{$R_V$}\\
\colhead{} & \colhead{(mag)} & \colhead{}}
\startdata
EW(Na~{\sc i}~D)\tablenotemark{a}                    & $\sim 0$--2 & \nodata \\
                                                                            & $1.26 \pm 0.29$ & \nodata \\
                                                                            & $1.08 \pm 0.25$ &  \nodata \\
EW(DIB $\lambda$5780)\tablenotemark{b}        & $\lesssim 1.06$ & \nodata \\
$E(V-R)$\tablenotemark{c}                                  & $1.07 \pm 1.35$ & 3.1 \\                                  
$E(B-V)$, $E(V-R)$, $E(V-I)$\tablenotemark{a} & 1.3--1.9 & 3.3--6.8 \\
$E(B-V)$ and $E(V-R)$ only\tablenotemark{a}   & 1.3--1.7 & 4.3 \\
                                                                             & 1.0--1.2 & 3.1 \\
\enddata
\tablenotetext{a}{Following \citet{Stritzinger+2018}.}
\tablenotetext{b}{Following \citet{Phillips+2013}.}
\tablenotetext{c}{Following \citet{Drout+2011}.}
\end{deluxetable}

The extinction to SN 2017ein, as it turns out, is not well determined from our data.
We first attempted to estimate the visual extinction from the SN spectra. 
\citet{Stritzinger+2018} presented for SESNe a correlation between the equivalent width (EW) of the Na~{\sc i}~D feature and 
host-galaxy $A_V$.
From the blended Na~{\sc i}~D feature in the June 2 (day $-6.3$) spectrum, arising entirely from the host galaxy, we measured EW= $1.61 \pm 0.06$\,\AA.
From the MMT spectrum, we measured from the resolved Na~{\sc i}~D1 and D2 lines an EW of $0.67 \pm 0.04$\,\AA\ and
$0.72 \pm 0.01$\,\AA, respectively. The total EW of the feature from this spectrum is then $1.39 \pm 0.05$\,\AA.
Referring to \citet[][ their Figure 17]{Stritzinger+2018}, we then can see that $A_V$ is somewhere in the range of 0 to 
$\sim 2$\,mag. Their best fit to the relation, $A_V = 0.78(\pm 0.15) \times$ EW(Na~{\sc i}~D), results in host 
$A_V = 1.26 \pm 0.29$ and $A_V=1.08 \pm 0.25$\,mag from the blended feature in the Lick spectrum and
the sum of the resolved lines in the MMT spectrum, respectively.

Unfortunately, the DIB $\lambda$5780 feature is not particularly distinct in the MMT spectrum.
Following \citet{Phillips+2013}, we fit a Gaussian of 2.1 \AA\ full width at half-maximum intensity (FWHM) to
the spectrum and placed a 3$\sigma$ upper limit of 203 m\AA\ to its equivalent width, which, from their Equation 6
(which has a 50\% systematic uncertainty), corresponds to $A_V \lesssim 1.06$ mag.

We can also estimate the amount of extinction to SN 2017ein through a color-curve comparison. 
\citet{Drout+2011}, from their systematic study of SN Ib/c light curves, found that the intrinsic color, 
$(V-R)_0 = 0.26 \pm 0.06$\,mag 
on $V$(10\,d), 10 days past $V$(max) (see also \citealt{Bianco+2014} and \citealt{Dessart+2016}). 
For SN 2017ein, we estimate that $V$(10\,d) was about JD 2,457,923.1, and on that
day, $V-R=0.49 \pm 0.27$\,mag. 
The Galactic foreground contribution to $E(V-R)$ is quite small, 0.012\,mag \citep{Schlafly+2011}.
This would indicate that the host contribution to $E(V-R)$ is $0.22 \pm 0.28$\,mag for SN 2017ein, where the uncertainty includes 
both the measurement uncertainty in the
SN 2017ein $V-R$ color from KAIT and the uncertainty in the \citet{Drout+2011} intrinsic color. 
Assuming a \citet{Fitzpatrick1999} reddening law and $R_V=3.1$, this corresponds to $A_V=1.07 \pm 1.35$ mag.

We compared the SN 2017ein $B-V$ curve, corrected for the low Galactic foreground reddening (0.019\,mag), with that of
SN 2007gr, for which the Galactic reddening is $E(B-V)=0.055$\,mag \citep[][ via \citealt{Schlafly+2011}]{Chen+2014} and the 
host reddening is assumed to be quite low \citep{Hunter+2009,Drout+2011,Chen+2014}; see Figure~\ref{figcolor}. Additionally, 
\citet{Stritzinger+2018}, based on their sample from the Carnegie SN Project (CSP), provided color templates, after 
correction for Galactic foreground reddening, for SESNe, specifically for SNe Ic, from 0\,d to +20\,d relative to 
$V$(max). The relevant template to apply here is for $B-V$, which we show compared to the color curve for SN 2017ein in 
Figure~\ref{figcolor}.
(Two other templates from \citealt{Stritzinger+2018}, $V-r$, and $V-i$, might have been applicable here; however, it is not
readily evident how to transform accurately for SNe Ic between the SDSS $r$ and $i$ bands used by the CSP and 
the Johnson-Cousins $R_C$ and $I_C$ bands which we used for this study.)
Both the template and the SN 2007gr $B-V$ color curve imply that $E(B-V) =0.34 \pm 0.07$\,mag for SN 2017ein.
From a minimum $\chi^2$ fitting of
the SN 2017ein color curves to the corresponding SN 2007gr ones, we estimate that $E(V-R)=0.24 \pm 0.06$ and 
$E(V-I)=0.57 \pm 0.06$\,mag for SN 2017ein (the Galactic foreground contribution to $E(V-I)$ is 0.026\,mag).
The value of $E(V-R)$ for the entire color curve is consistent with that inferred from the \citet{Drout+2011} fiducial value at 
$V$(10\,d).

We performed an analysis of the three color excesses in a manner similar to that of \citet{Stritzinger+2018}, assuming a 
\citet{Fitzpatrick1999} reddening law. We fit the values of host $A_V$ and $R_V$ that were best constrained by all
three excesses.
We found a rather large range in both parameters, $A_V=1.3$--1.9\,mag for $R_V=3.3$--6.8.
Immediately, these imply that the visual extinction to SN 2017ein is appreciable, consistent with the large range in $A_V$ inferred
from the Na~{\sc i} D feature strength, and that the dust is dissimilar from the diffuse Galactic component (for which $R_V=3.1$).
\citet{Stritzinger+2018} have indeed argued that $R_V=4.3$ is typical for SNe Ic, consistent with their general location in dusty
environments presumably comprised of larger dust grains (we note, however, that this inference is based on only one of the 
events in their sample, which had more extreme reddening). Values of $R_V$  larger than 3.1 are typical for massive star-forming
regions; for example, \citet{Smith2002} found that the local $R_V=4.8$ for clouds in the Carina Nebula. This effect likely arises 
from the
strong ultraviolet radiation from young O-type stars which destroys some of the smallest grains, leading to a flatter reddening law.

The high values of $A_V$ and $R_V$ are most strongly driven in our analysis by the value of $E(V-I)$. 
If we relax our analysis and consider only $E(B-V)$ and $E(V-R)$, and if we further assume that $R_V=4.3$ (see above), then 
$A_V$ would be in the range of 1.3--1.7\,mag, consistent with the range in $A_V$ when including all three color excesses.
If we, however, assume that $R_V=3.1$, similar to Galactic diffuse interstellar dust, then $A_V$ would be distinctly lower, 
1.0--1.2\,mag.
For the sake of discussion below, specifically when we analyze the nature of the SN progenitor in Section~\ref{progenitor}, we 
will consider all three ranges in extinction and their implications.

We summarize all of the host-galaxy extinction estimates for SN 2017ein in Table~\ref{tabext}. If a value for $R_V$ is 
either estimated or assumed, it is indicated in the table as well.

\subsection{Distance to SN 2017ein}

Another quantity for SN 2017ein that is not well known is its distance, $D$. Several estimates exist for the distance to the host 
galaxy, NGC 3938. \citet{Poznanski+2009}, based on assuming that SNe II-P are standardizable candles, estimated
that the distance modulus to SN 2005ay is $\mu = 31.27 \pm 0.13$\,mag ($D = 17.9$\,Mpc). \citet{Rodriguez+2014}, invoking a 
similar, color-based standardization of the absolute brightness of SNe II-P, found that $\mu = 31.75 \pm 0.24$\,mag (22.4\,Mpc)
for their method in the $V$ band and $31.70 \pm 0.23$\,mag (21.9\,Mpc) in $I$.
Several early Tully-Fisher estimates \citep{Bottinelli+1984,Bottinelli+1986}
resulted in far shorter distances, with $\mu \approx 28.7 \pm 0.7$\,mag ($D \approx 5.7$\,Mpc), 
although \citet{Tully1988} lists $\mu = 31.15 \pm 0.40$\,mag (17.0 Mpc)\footnote{See also the Extragalactic Distance
Database, http://edd.ifa.hawaii.edu/.}.

We therefore consider hereafter a range in possible distance moduli to SN 2017ein of 31.15 to 31.75\,mag ($D=17.0$ to 
22.4\,Mpc). Given this distance range and the inferred SN extinction, above, for SN 2017ein, $M_V ({\rm max})\approx -16.9$ to
$-17.5$\,mag, which is consistent with $-17.2$\,mag for SN 2007gr \citep{Hunter+2009}, but less luminous than $-18.0$\,mag 
for SN 2004aw \citep{Taubenberger+2006}.

\subsection{Quasi-Bolometric Light Curve\label{bolo}}

\begin{figure}
\plotone{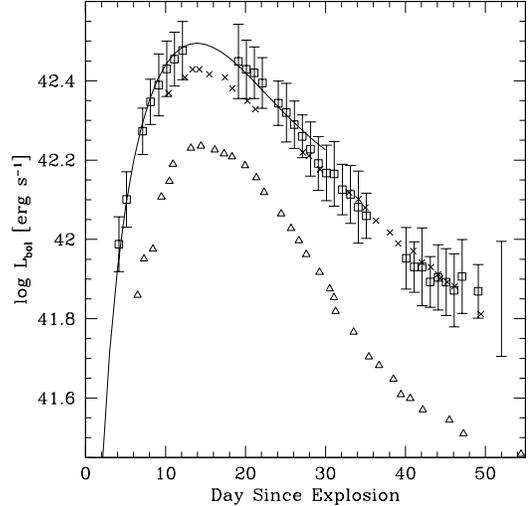}
\caption{Quasi-bolometric light curve of SN 2017ein (open squares), assuming the bolometric corrections for SESNe from 
\citet{Lyman+2014}. The uncertainties shown with each data point arise primarily from the photometric measurements and from 
the uncertainties in the bolometric corrections. An error bar is also given, representing the additional uncertainty in both the 
reddening and the distance. For comparison we show the bolometric light curves for SN 2007gr 
\citep[][ open triangles]{Hunter+2009,Chen+2014} 
and SN 2004aw \citep[][ crosses]{Taubenberger+2006}. Additionally, we display the mean best-fit \citet{Arnett1982} 
semi-analytical model (solid curve), powered by radioactive decay of $^{56}$Ni and $^{56}$Co.\label{figbolo}}
\end{figure}

\begin{deluxetable*}{ccccc}
\tablecaption{Comparison of Light-Curve-Derived Parameters for SN 2017ein\label{tabparams}}
\tablecolumns{5}
\tablewidth{0pt}
\tablehead{
\colhead{SN} & \colhead{$M (^{56}{\rm Ni})$} & $M_{\rm ej}$ & $E_{\rm K}$ & \colhead{Source} \\
\colhead{} & \colhead{($M_{\odot}$)} & \colhead{($M_{\odot}$)} & \colhead{($10^{51}$ erg)} & \colhead{}}
\startdata
2017ein & 0.10--0.17 & 0.96--1.76 & {0.54--0.99} & This work\tablenotemark{a} \\
$\cdots$ & 0.05--0.2 & $1.2^{+0.3}_{-0.2}$ & $0.7^{+0.2}_{-0.1}$ & This work\tablenotemark{b} \\
2007gr & $0.061 \pm 0.014$ & \nodata & \nodata & \citet{Chen+2014}\tablenotemark{a} \\
$\cdots$ & $0.076 \pm 0.020$ & 2.0--3.5 & 1--4 &  \citet{Valenti+2008,Hunter+2009}\tablenotemark{a} \\
$\cdots$ & \nodata & $\sim 1.0$ & \nodata & \citet{Mazzali+2010}\tablenotemark{c} \\
$\cdots$ & $0.07^{+0.01}_{-0.01}$ & $1.2^{+0.6}_{-0.4}$ & $0.8^{+0.3}_{-0.3}$ & \citet{Drout+2011}\tablenotemark{b} \\
2004aw & 0.25--0.35 & 3.5--8.0 & 3.5--9.0 & \citet{Taubenberger+2006}\tablenotemark{a}; \citet{Mazzali+2017}\tablenotemark{c} \\
$\cdots$ & $0.27^{+0.05}_{-0.05}$ & $4.5^{+2.1}_{-1.3}$ & $2.8^{+1.3}_{-0.8}$ & \citet{Drout+2011}\tablenotemark{b} \\
SN Ic sample & $0.33 \pm 0.07$ & $5.75 \pm 2.09$ & $1.75 \pm 0.24$ & \citet{Taddia+2015}\tablenotemark{a} \\
$\cdots$ & $0.24 \pm 0.15$ & $1.7^{+1.4}_{-0.9}$ & $1.0^{+0.9}_{-0.5}$ & \citet{Drout+2011}\tablenotemark{b} \\
\enddata
\tablenotetext{a}{From the quasi-bolometric light curve.}
\tablenotetext{b}{From multiband light curves.}
\tablenotetext{c}{From analysis of nebular spectra.}
\end{deluxetable*}

We constructed an early-time bolometric light curve of SN 2017ein from our observed $BVR_CI_C$ light curves. 
Given the relative sparseness in the photometric coverage and the large ranges in both distance and reddening, high precision
was not a particular concern. For that reason, rather than performing our own detailed blackbody fitting of each set of photometric 
points, we utilized the relations for SESNe from \citet{Lyman+2014} for 
estimating bolometric corrections to the absolute $B$ and $V$ light curves based on the reddening-corrected colors, in order
to generate the SN 2017ein bolometric light curve. We show this curve in Figure~\ref{figbolo}. Although there are uncertainties
in both the photometry and the bolometric corrections, these are dwarfed by the overall uncertainties in both the distance and the
extinction to the SN.

For comparison we also show in Figure~\ref{figbolo} the bolometric light curves of SN 2007gr \citep{Hunter+2009,Chen+2014}
and SN 2004aw \citep{Taubenberger+2006}.
The bolometric luminosity of SN 2017ein is consistent at peak, to within the large uncertainties, with that of SN 2007gr; however,
overall, the former is generally more luminous than the latter. 
SN 2004aw, which was more luminous than SN 2007gr, is in agreement with SN 2017ein, to within the 
uncertainties.

We have modeled the bolometric luminosity via a semi-analytical light curve, 
following the method of 
\citet[][ see also \citealt{Valenti+2008,Cano2013,Taddia+2015,Lyman+2016,Prentice+2016,Arnett+2017}]{Arnett1982}.
The best-fit model to the bolometric luminosity implies that the maximum bolometric luminosity, $L_{\rm bol}$(max), is
(1.7--3.4) $\times 10^{42}$\,erg\,s$^{-1}$, and the maximum occurred around day 14 since 
explosion. We show the mean of the best-fit model in Figure~\ref{figbolo}.

The \citet{Arnett1982} model (see, e.g., Equations A1 and A2 of \citealt{Valenti+2008}; Equations 1 and 2 of \citealt{Cano2013})
fits two parameters, the nickel mass $M(^{56}{\rm Ni})$ and the diffusion timescale
$\tau_{\rm m}$, where
$\tau_{\rm m}^2 = (C \kappa_{\rm opt} M_{\rm ej}) / (\beta c v_{\rm sc})$. 
In this expression, $M_{\rm ej}$ is the SN ejecta mass, $c$ is the speed of light, $\beta$ is a constant of integration 
$\sim 13.8$ \citep{Arnett1982}, $v_{\rm sc}$ is the scale velocity, $\kappa_{\rm opt}$ is the optical opacity (often 
assumed to be 0.07\,cm$^2$\,g$^{-1}$), and $C$ is a constant of proportionality. 
Note that the $M_{\rm ej}$ estimate is just for the mass involved in the diffusion of SN light; if nonionized ejecta mass also exists,
the value of $M_{\rm ej}$ could be larger \citep{Wheeler+2015}.
The quantity $v_{\rm sc}$ is generally assumed to be the photospheric velocity ($v_{\rm ph}$) at maximum light. 
Note that, for example, \citet{Lyman+2016} assumed that $C=2$, whereas \citet{Chatzopoulos+2012} assumed $C=10/3$ (since,
generally, the mean photospheric velocity is adopted, rather than the peak velocity, at optical depth 1/3).
The kinetic energy of the ejecta is $E_K= (3/10)\,M_{\rm ej} v_{\rm sc}^2$.
From our fitting, the value of $\tau_{\rm m}=11.6$\,d.
Based on our measurement of velocities from the Fe~{\sc ii} lines (assumed to be nearly photospheric) in the observed spectra, 
we have set $v_{\rm ph}=9700$\,km\,s$^{-1}$.

The results of the modeling are that $M (^{56}{\rm Ni})$ is in the approximate range 0.09--0.18 $M_{\odot}$.
If $\kappa_{\rm opt}=0.07$\,cm$^2$\,g$^{-1}$, then for SN 2017ein, $M_{\rm ej} \approx 1.45\ M_{\odot}$ and, thus,
$E_K \approx 8.1 \times 10^{50}$\,erg.
For comparison with SN 2007gr, 
$M (^{56}{\rm Ni}) \approx 0.06\ M_{\odot}$ \citep{Chen+2014} to $\sim 0.08\ M_{\odot}$ 
\citep{Valenti+2008,Hunter+2009}, 
and \citet{Hunter+2009} estimated that $M_{\rm ej} \approx 2.0$--3.5 $M_{\odot}$ and 
$E_K \approx$ (1--4) $\times 10^{51}$\,erg.
From nebular spectra of SN 2007gr, \citet{Mazzali+2010} estimated a lower $M_{\rm ej} \approx 1\ M_{\odot}$.
In contrast, \citet{Taubenberger+2006} found for SN 2004aw significantly higher values of 
$M (^{56}{\rm Ni})=0.25$--0.35 $M_{\odot}$, $M_{\rm ej}=3.5$--8.0 $M_{\odot}$, and $E_K =$ (3.5--9.0) $\times 10^{51}$\,erg
(see also \citealt{Mazzali+2017}).
We note that \citet{Taddia+2015}, from their sample of SNe Ic, estimated mean values of 
$M (^{56}{\rm Ni})=0.33$\,M$_{\odot}$, $E_K=1.75 \times 10^{51}$\,erg, and large $M_{\rm ej}=5.75\ M_{\odot}$.
We summarize this comparison in Table~\ref{tabparams}.

We can also analyze these parameters, $M(^{56}{\rm Ni})$, $M_{\rm ej}$, and $E_K$, from the observed light curve,
following \citet{Drout+2011}.
From the simple analytic model we constructed for the $R$-band light curve (see Section~\ref{secphot}), we found that 
$M_R({\rm max}) \approx -17.1$ to $-18.6$\,mag for SN 2017ein, given the distance and extinction range. From this model,
the quantity $\Delta m_{15}$, the decrease in brightness between maximum and 15 days post-maximum, is 0.87\,mag.
This $\Delta m_{15}$ value is consistent with that of SN 2007gr, to within the uncertainties \citep{Drout+2011}.
From $M_R({\rm max})$ and $\Delta m_{15}$, we found that $M(^{56}{\rm Ni})\approx 0.05$--0.2 $M_{\odot}$, which agrees
well with that found from the bolometric curve. From the simple model, the characteristic width of the light curve is
$\tau_c = 10 \pm 1$ d (again, consistent with SN 2007gr; \citealt{Drout+2011}). For 
$v_{\rm ph}=9700$\,km\,s$^{-1}$, we can then
infer that $M_{\rm ej} \approx 1.2\ M_{\odot}$ and $E_K \approx 7 \times 10^{50}$\,erg for SN 2017ein. Again, these are 
completely consistent both with what we found (above) for the bolometric light-curve analysis and with SN 2007gr.
We note that \citet{Drout+2011} found much higher values for these various parameters for SN 2004aw: 
$\tau_c \approx 19$, $M (^{56}{\rm Ni})\approx 0.27\ M_{\odot}$, $M_{\rm ej} \approx 4.5\ M_{\odot}$, and 
$E_K \approx 2.8 \times 10^{51}$\,erg.

\section{The Supernova Progenitor}

\subsection{Progenitor Identification}\label{prog}

\begin{figure}
\centering
\includegraphics[width=0.33\textwidth]{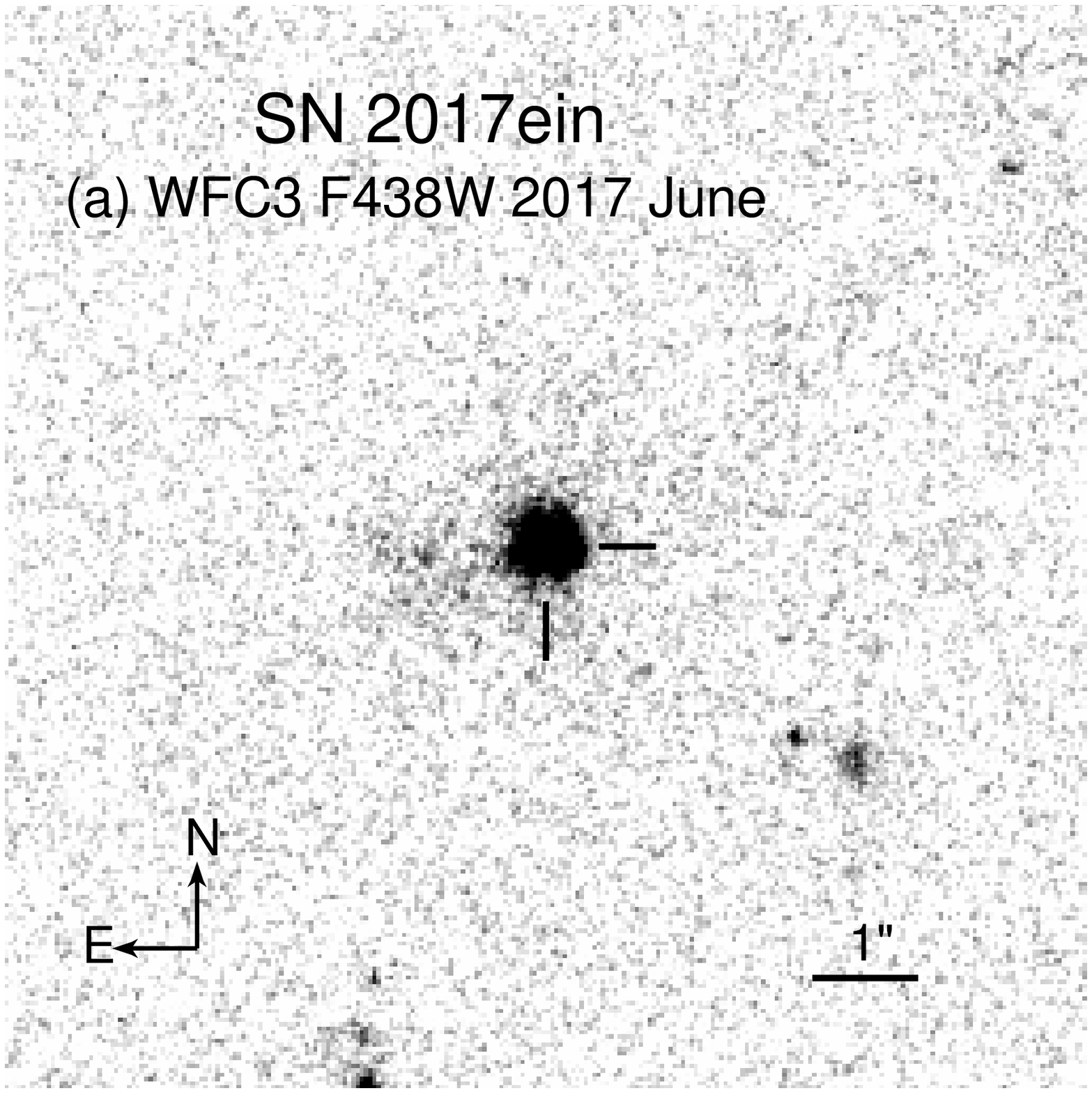}
\includegraphics[width=0.33\textwidth]{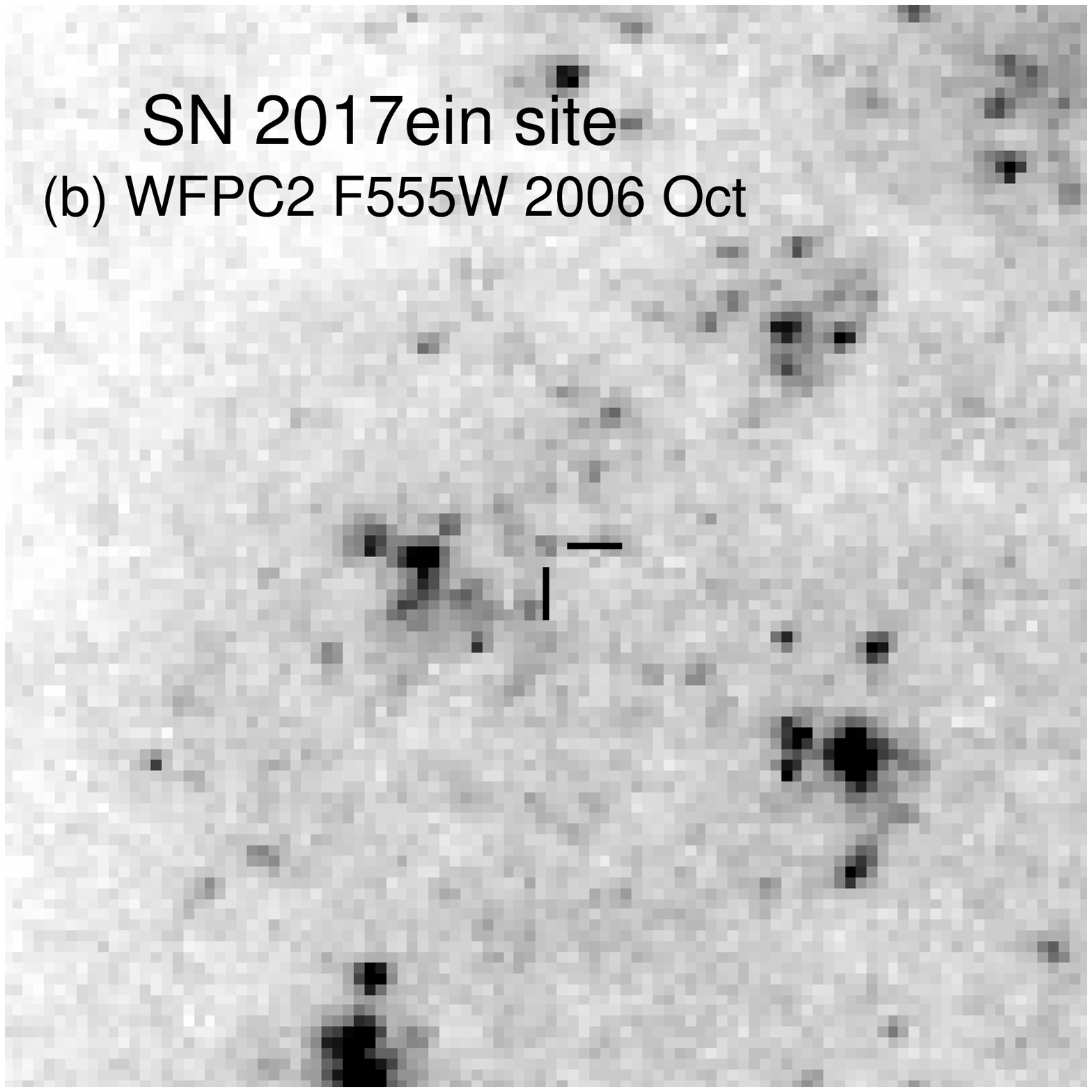}
\includegraphics[width=0.33\textwidth]{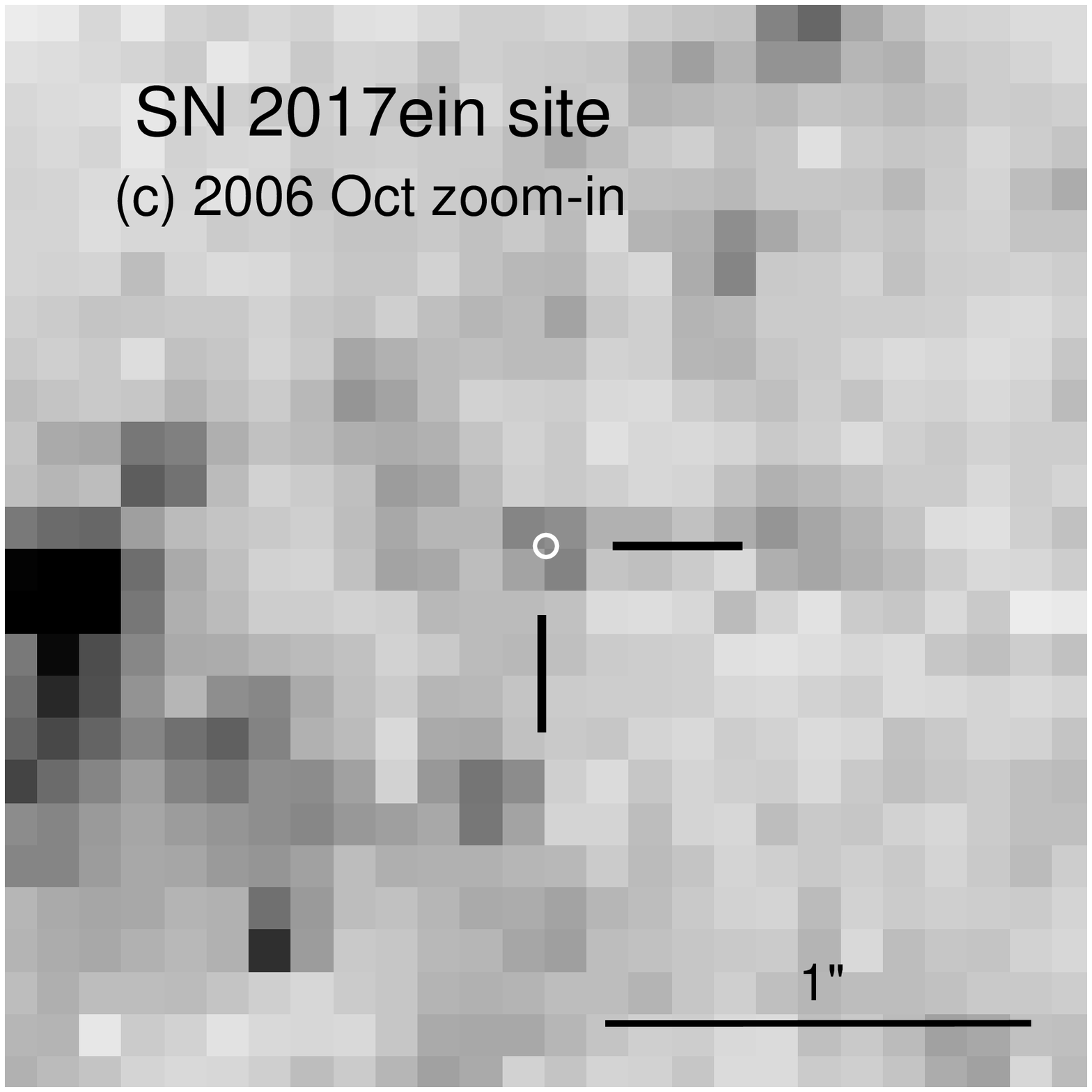}
\caption{(a) A portion of the WFC3/UVIS image of SN 2017ein obtained in F438W on 2017 June 12. The position of the SN is
indicated with tickmarks. (b) A portion of the WFPC2 image of the host galaxy, NGC 3938, in F555W on 2006 October
20, at the same orientation and scale as panel (a). (c) A zoom-in of panel (b), with the SN position indicated by a 
white circle with radius equal to the 0.26 pixel rms astrometric uncertainty. The location of the SN is also indicated in all three 
panels with tickmarks. We consider the identified object to be the candidate progenitor of SN 2017ein. 
North is up, east is to the left in the figure.\label{figsn}}
\end{figure}

To possibly identify a  progenitor candidate in the 2006 {\sl HST\/} images,
we astrometrically registered the 2006 WFPC2 F555W image mosaic to the 2017 WFC3 image mosaic. 
Using 27 fiducial stars, we were able to register the images to 0.26 WFPC2/WF pixel  [$1\sigma$ root-mean square (rms); 
26 milliarcsec].
As a result, SN 2017ein would be at (798.55, 1766.58) in the WFPC2 image mosaic. An object can be seen at (798.33, 1766.54).
This is a difference of 0.22 pixel, which is within the rms uncertainty in the astrometry. 
We therefore consider this object to be a candidate for the progenitor of SN 2017ein.
Note that we first preliminarily identified this object in \citet{VanDyk+2017}. 
We show the WFC3 and WFPC2 images to the same scale and orientation in Figure~\ref{figsn}.

We have analyzed this object further by performing PSF-fitting photometry of the {\sl HST\/} WFPC2 images using Dolphot 
\citep{Dolphin2000,Dolphin2016}. 
Prior to this step, we processed the individual WFPC2 frames with AstroDrizzle, to attempt to flag cosmic-ray hits.
We set the Dolphot parameters FitSky=3 and RAper=8, using the TinyTim PSFs provided 
with Dolphot and setting the parameters InterpPSFlib and WFPC2useCTE to ``true.'' The SN site is contained in the WFPC2 chip 
2. We find $24.56 \pm 0.11$ and $24.58 \pm 0.17$ VEGAMAG in F555W and F814W, respectively.
Dolphot also outputs ``object type''=``1'' for the source, which indicates that the routine considers it to be a ``good star.'' 
Other indicators also point to a star-like character: the sharpness and crowding parameters in each band are 0.002 and $-0.007$,
and 0.030 and 0.076, respectively.
The 
signal-to-noise ratio (SNR) in each band for the star is appreciable, $\sim 10$ and $\sim 6$ in F555W and F814W, respectively.
The output photometric quality flag from Dolphot is also ``0'' for both bands, meaning (from the Dolphot documentation) that the 
star was recovered very well in the image data.
The observed color of the source is F555W$-$F814W$=-0.02 \pm 0.20$\,mag.

\subsection{Metallicity of the Supernova Site}

We can estimate the metallicity at the SN 2017ein site from the oxygen abundance gradient for NGC 3938 presented by 
\citet{Pilyugin+2014}. Oxygen abundance is often used as a proxy for metallicity.
In that study the central oxygen abundance is $12+\log({\rm O/H}) = 8.79 \pm 0.05$ and the gradient is 
$-0.0413 \pm 0.0066$ dex kpc$^{-1}$.
We deprojected an image of the host galaxy from the Digitized Sky Survey, assuming the position angle ($29\arcdeg$) and 
inclination ($18\arcdeg$) from \citet{Jarrett+2003}, and calculate that the SN is $44\arcsec$ from the host nucleus.
Given the range in host-galaxy distance (above), this corresponds to a $\sim 3.7$--4.8\,kpc nuclear offset, and with this range 
we estimate, from the published abundance gradient, that $12+\log({\rm O/H}) \approx 8.59$--8.64. 

Nebular H$\alpha$ and [N~{\sc ii}]$\lambda$6583 emission are also seen in the MMT spectrum from June 24. Although not an 
ideal strong-line indicator, the ratio of the line fluxes, [N~{\sc ii}]$\lambda$6583/H$\alpha$, known as the $N_2$ index 
\citep{Pettini+2004}, provides an estimate of metallicity. Dereddening the spectrum assuming both $A_V=1$\,mag (with 
$R_V=3.1$), we obtain $N_2 = -0.30$. If the extinction in the SN environment were as high as $A_V=2$\,mag, the index 
varies slightly, to $-0.31$. Following the calibration of this index by \citet{Marino+2013}, this would imply that 
$12+\log({\rm O/H}) \approx 8.60$.

If we adopt the oxygen abundance for the Sun as $12+\log({\rm O/H}) \approx 8.69$ \citep{Asplund+2009}, we can  
infer that the metallicity at the SN site is somewhat subsolar (i.e., [Fe/H] $\approx -0.10$).

\subsection{The Nature of the Progenitor Candidate\label{progenitor}}

\begin{figure}
\plotone{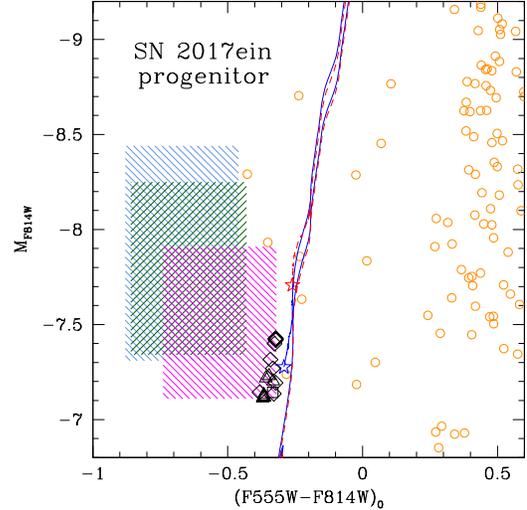}
\caption{Color-magnitude diagram, in {\sl HST\/} WFPC2 bands, showing three possible ranges in loci for the SN 2017ein 
progenitor candidate as hashed regions: (1) for  $A_V=1.3$--1.9\,mag and $R_V=3.3$--6.8, light blue; (2) for 
$A_V=1.3$--1.7\,mag fixed at $R_V=4.3$, dark green; and, (3) for $A_V=1.0$--1.2\,mag fixed at $R_V=3.1$, magenta. Also 
included in these ranges is the uncertainty in the distance to the host galaxy. 
Shown for comparison are single-star MIST evolutionary tracks
\citep{Paxton+2011,Paxton+2013,Paxton+2015,Choi+2016} for 48 $M_{\odot}$ (red dashed curve) and 49 $M_{\odot}$ 
(blue solid curve); the endpoint of each 
track is indicated with a star. Additionally, we show the endpoints of model binary systems from BPASS v2.1 
\citep{Eldridge+2017} at three possible metallicities, 
$Z=0.020$ (open square), $Z=0.014$ (open triangles), and $Z=0.010$ (open diamonds).
The colors and magnitudes of stellar clusters in NGC 628 (M74) identified by \citet{Adamo+2017} are represented by orange 
open circles. See text for details.\label{figcmd}}
\end{figure}

Assuming that the extinction to the SN we have estimated applies to the progenitor candidate as well, and 
given the assumed range in distance, we find 
that the object is consistent with being quite luminous and blue.
(Note that we are only considering here interstellar extinction, which provides a lower limit on the total extinction in
the presence of any circumstellar extinction destroyed by the SN breakout; although not typically expected for hot blue stars,
some binary WC stars are surrounded by dust formed by colliding winds; e.g., \citealt{Williams2008}.)
We show in Figure~\ref{figcmd} three possible ranges of loci for the object in a color-magnitude diagram (CMD), based on our
assumptions about the amount of reddening, as discussed in Section~\ref{extinction}.

We compared these loci in Figure~\ref{figcmd} with single-star evolutionary tracks from the Modules for Experiments in 
Stellar Astrophysics (MESA) 
Isochrones and Stellar Tracks  \citep[MIST;][]{Paxton+2011,Paxton+2013,Paxton+2015,Choi+2016} v1.1 with rotation at 
$v_{\rm rot}/v_{\rm critical}=0.4$, solar-scaled abundances \citep[assuming $Z=0.0142$;][]{Asplund+2009}, and 
subsolar metallicity [Fe/H]$=-0.10$. The tracks have been interpolated to the WFPC2 
VEGAMAG system by the MIST online interface\footnote{http://waps.cfa.harvard.edu/MIST/.}.
Among the tracks, we find that the candidate's locus, assuming the lowest reddening, is roughly consistent with, although 
somewhat blueward of, the terminus of a star with initial mass 48--49 $M_{\odot}$.
The net effect of rotation is to lead to a bigger He core mass, and thus more luminous progenitor, for the same initial mass.  
If the star was not a rapid rotator, then the implied initial mass (if single) would be even higher.
We note that some evolutionary models for single stars with these initial masses at solar metallicity experience ``failed'' 
explosions, with the cores collapsing directly to black holes \citep[e.g.,][]{Sukhbold+2016}.

Additionally, we compared in Figure~\ref{figcmd} with the endpoints of binary-star models from BPASS v2.1 
\citep{Eldridge+2017} at two different 
metallicities, solar and somewhat subsolar (note that for BPASS, $Z=0.020$ is considered solar metallicity).
What is shown in the figure is the combined light of both the primary and the secondary in each of the binary models.
(Here we have assumed that the F555W and F814W bandpasses used for BPASS are generic enough to apply for WFPC2.)
We have considered only systems for which the primary's
surface H mass fraction $=0$, surface He mass fraction $\le 0.27$, and the ratio of the mass of remaining He to the total 
ejecta mass $\le 0.1$ at the model endpoint. 
These criteria should be sufficient to approximate a He-stripped SN Ic progenitor (J.~J.~Eldridge, private communication). 
The systems allowed by the color and luminosity range for the progenitor, again assuming the lowest reddening, are to the 
lower-right corner of this color-luminosity range.
At $Z=0.020$ the lone system has a 60 $M_{\odot}$ primary, a mass ratio of the two components $q=0.9$ (i.e., the system 
consists of two stars of nearly the same mass), and initial orbital period $\log P ({\rm day})=1.0$.
The primary of this system ejects $M_{\rm ej}\approx 10\ M_{\odot}$ and leaves behind a remnant with 
$\sim 6\ M_{\odot}$.
At $Z=0.014$ there is one system with these same parameters, as well as systems with a 80 $M_{\odot}$ primary, $q=0.7$, and
a range of $\log P=0.8$ to 2.0.
These primaries eject $M_{\rm ej}\approx 6\ M_{\odot}$ and have remnants with $\sim 7$--15 $M_{\odot}$.
At $Z=0.010$ the systems have primaries which, similarly, range from 60 to 80 $M_{\odot}$, with $q=0.7$--0.8, and 
$\log P=1$--2.8. Ejecta masses are in the range 5--8 $M_{\odot}$, and remnant masses are $\sim 8$--12 $M_{\odot}$.
The remnants for all of these systems are most likely to be black holes.

It may seem counterintuitive that the model binary systems should possess a primary that is more massive than the 
single-star models, with both binary and single-star models having essentially the same observed luminosity in F814W at their 
endpoints.
However, it should be remembered that what we are showing in Figure~\ref{figcmd} is a CMD and not
a Hertzsprung-Russell diagram (HRD). For all of the binary models considered here, the mass transfer is generally
nonconservative: The more massive primaries in the models all lose substantial amounts ($\gtrsim 75$\%) of their initial 
mass, while the secondaries increase little in
mass as the systems evolve. On a HRD the primaries end up being less luminous bolometrically, 
although far hotter ($T_{\rm eff}>10^5$ K),
than they were on the zero-age main sequence; the secondaries generally increase monotonically in luminosity, while 
changing little in effective temperature, ending up more luminous than the primaries. In fact, the model primaries in the binary 
systems end up less luminous than the endpoints of
the single-star models. In the CMD shown in the figure, the secondaries all dominate the observed light of the systems at their 
endpoints; as an additional effect, the hot, stripped primaries contribute comparatively far less light at F814W.

It appears that the two highest-reddening scenarios, $A_V=1.3$--1.9\,mag for $R_V=3.3$--6.8 and $A_V=1.3$--1.7\,mag at 
$R_V=4.3$, imply that the object would be too blue and possibly too luminous for single stars or binary star systems. The 
lowest-reddening scenario also may be just on the verge of being too blue and luminous.

In addition, we cannot rule out that the progenitor candidate is actually a compact cluster.
At the distance of NGC 3938, such a cluster could be unresolved.
To that end, we considered stellar clusters in a galaxy similar to NGC 3938, those in NGC 628 (M74; also a nearly face-on Sc 
galaxy, but closer to us at 10.19 Mpc; \citealt{Jang+2014}), which
\citet{Adamo+2017} studied (see also \citealt{Grasha+2015}). We considered the ``default,'' or ``reference'' deterministic catalog, for which the cluster 
photometry is an averaged aperture correction method, and for which Milky Way extinction and Padova stellar evolutionary 
tracks for age-dating are adopted\footnote{Available at
\url{https://archive.stsci.edu/prepds/legus/dataproducts-public.html};
\dataset[doi:10.17909/T9J01Z]{https://doi.org/10.17909/T9J01Z}.
}.
We selected from the catalog only those clusters that are symmetric, compact ones and include these for comparison in 
Figure~\ref{figcmd}. 
(Here we neglected differences in the F555W and F814W bandpasses between WFC3/UVIS and WFPC2.)
We have dereddened the clusters, using the ``best $E(B-V)$'' in the catalog and assuming $R_V=3.1$. 
Note that $\sim 5$ of these clusters are also roughly consistent with the possible loci for the progenitor. 
These particular clusters are all quite blue and young, with estimated ages $\lesssim 4$ Myr and masses 
$\lesssim 2000\ M_{\odot}$.
Based on the MIST tracks, the turnoff masses for such young clusters are $\gtrsim 65\ M_{\odot}$.

\subsection{Constraints from the Stellar Environment}

\begin{figure}
\centering
\includegraphics[angle=270,width=0.47\textwidth]{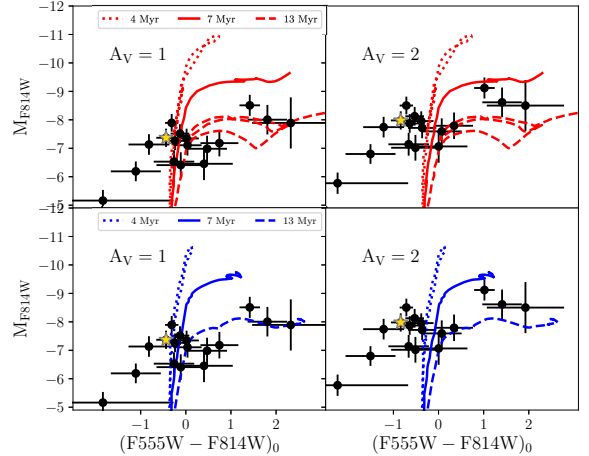}
\caption{Color-magnitude diagrams of the stellar environment (points) within $\sim 100$\,pc of SN 2017ein, based on the
{\sl HST\/} photometry, corrected by different assumptions for total visual extinction and analyzed with different model
isochrones. The uncertainty in the distance to NGC 3938 is included in the uncertainties in absolute magnitude.
The left panels assume $A_V=1$\,mag toward the environment; the right panels assume
$A_V=2$\,mag. The top panels show PARSEC-COLIBRI isochrones \citep[][ red curves]{Marigo+2017}, and the bottom panels 
show MIST isochrones \citep[][ blue curves]{Choi+2016}, at $Z=0.01$ with ages 4, 7, and 13\,Myr. The progenitor candidate is
indicated by the yellow (light) star and has been reddening-corrected by the same amount as the overall 
environment.\label{figenv}}
\end{figure}

We can also potentially constrain the initial mass of the SN 2017ein progenitor by estimating the age of the stellar 
population, or populations, in its immediate environment. Here we consider a radius from the SN position for the environment
of $\sim 100$\,pc, a typical size for an OB association, within which stars should generally be coeval with the progenitor. 
We had already extracted the photometry from the archival {\sl HST\/} WFPC2 images for the environment, as well as for
the progenitor (Section~\ref{prog}). All of these objects appear to be point-like, based on their Dolphot photometric properties,
discussed above. We show in Figure~\ref{figenv} a CMD displaying the objects in the environment, corrected by two different
assumptions of the total visual extinction, $A_V=1$ and 2\,mag, and adjusted by the assumed
distance (the uncertainty in the distance is included in the uncertainties in absolute magnitude). A \citet{Cardelli+1989} 
reddening law was assumed here with $R_V=3.1$.
We note, of course, that this is a simplification, since, over this size scale the extinction could be variable.
The photometry for the 
progenitor candidate is included in the figure, but corrected for reddening by the same amounts as the overall environment,
not as we have done in Section~\ref{progenitor}.

One can see a number of stellar objects in the SN 2017ein environment. One interesting inference to point out is that, although
the progenitor candidate may be among the bluest objects in the environment, it is not necessarily the most luminous.
We have compared the photometry for these objects with the PARSEC-COLIBRI model isochrones from \citet{Marigo+2017} and
MIST isochrones from \citet{Choi+2016}, both at a subsolar metallicity $Z=0.01$. 
Note that these models are all based on single-star evolutionary tracks.
We can approximately fit these populations, using both sets of models, with a mix of three different ages (4, 7, and 13\,Myr), if the
extinction is $A_V \approx 1$\,mag. The objects are poorly fit if $A_V$ were closer to 2\,mag, particularly for the objects with 
$({\rm F555W}$-${\rm F814W})_0 <0$\,mag, which are likely massive main-sequence stars.  We therefore consider this assumed
extinction to be an upper limit on the actual extinction.
Assuming the lower extinction, the youngest objects are generally quite blue. 
Three evolved cool supergiants appear consistent with the 13\,Myr isochrone, with one or two additional objects consistent with
the blue loop of the PARSEC-COLIBRI track at that age.

The corresponding turnoff masses for the PARSEC-COLIBRI isochrones are $\sim 57$, 28, and 16 $M_{\odot}$. These 
masses are quite similar for the MIST isochrones, at  $\sim 60$, 26, and 15 $M_{\odot}$.
That a SN has occurred recently in this environment would imply that it is a member of the younger, more massive population.
The inferred initial masses are consistent with those of the high-mass BPASS binary models that appear to agree with the locus 
of the progenitor candidate in the CMD. However, if the assumption of coevality breaks down (for example, dispersal through 
random motions), then it is possible that the SN progenitor was a member of one of the older, less massive populations, 
consistent with the expectation of a more moderate-mass binary progenitor.
Furthermore, \citet{Maund2017} showed that, within 100\,pc in the environments of SNe II-P, a mixture of populations at different
ages could be colocated, one of those populations being related to a given SN II-P progenitor. A similar situation might be in effect
for the SN 2017ein environment.

\section{Discussion}

We have found that SN 2017ein in NGC 3938 appears to be similar, both photometrically and spectroscopically, to SN 
2007gr during approximately its first month since explosion. We have also compared SN 2017ein to SN 2004aw, to which it bears
more similarity spectroscopically than photometrically.
Uncertainty exists in the value of the extinction to SN 2017ein. We have attempted to estimate the extinction from the strength
of the Na~{\sc i} D absorption in the spectra, as well as via comparisons of the object's colors with those of SN 2007gr and a 
$B-V$ color template for SNe Ic assembled by \citet{Stritzinger+2018}. We were only able to establish a range in $A_V$, 
together with a possible range in $R_V$. Given the uncertainties in the overall reddening and in the distance to the host 
galaxy, we found that, bolometrically, SN 2017ein is somewhat more luminous than SN 2007gr and could be as luminous as 
SN 2004aw. We estimated $M (^{56}{\rm Ni})$, $M_{\rm ej}$, and $E_K$, which are reasonably consistent with the estimates 
for previous SNe Ic.

We have also identified a progenitor candidate in archival {\sl HST\/} WFPC2 imaging. This potentially constitutes the first time 
that a progenitor candidate has ever been identified for a SN Ic. Based on the assumed ranges in both distance and extinction 
to the SN, we infer that the progenitor candidate is quite luminous and blue, possibly more so than we would have expected, 
based on models of their progenitors.  Comparison with both single-star evolutionary tracks and the
endpoints of binary-star models would imply that the progenitor is very massive, with initial mass $\sim 49\ M_{\odot}$, if 
single, and possibly as high as $80\ M_{\odot}$, if the primary is in a binary system. 
This is far more 
massive than what we infer to be the initial masses of the progenitors of the more common SNe II-P and also of SNe IIb and Ib 
(although the true nature of the sole known SN Ib progenitor, of iPTF13bvn, is still to be established).
Such a high mass would be consistent with what \citet{Maund+2016} concluded for the progenitor of SN 2007gr, that it had to 
be quite massive, $M_{\rm ZAMS} \approx 30\ M_{\odot}$, based on an analysis of that object's stellar environment. 
However, we note that
\citet{Mazzali+2010} tallied up $\sim 1\ M_{\odot}$ of ejecta for SN 2007gr from late-time spectra and concluded that the
progenitor was a star of relatively low initial mass ($\sim 15\ M_{\odot}$).

We also considered the possibility that the observed object is a compact cluster, rather than a star or stellar system. 
We compared the candidate to observations of compact clusters in a galaxy similar in characteristics to the SN host. We 
found that the progenitor candidate is roughly consistent with among the bluest and youngest clusters, with very high turnoff 
masses, which would, again, imply a high-mass progenitor for SN 2017ein. 
If it were a very young, blue cluster, we might expect interstellar H$\alpha$ emission at the site. 
We indeed detect such emission, particularly in the early-time MMT spectrum, and we estimate from 
the observed flux in the H$\alpha$ line, the H$\alpha$ luminosity to be $L_{{\rm H}\alpha}\approx 6.7 \times 10^{49}$ to 
$2.5 \times 10^{50}$ erg s$^{-1}$, given the range of assumed reddening and distance. Assuming the number of Lyman 
continuum photons $N_{\rm LyC}=7.25 \times 10^{11}\ L_{{\rm H}\alpha}$, and adopting the stellar parameters from 
\citet{Martins+2005}, this is the equivalent of 1--4 O3~I stars, or 4--15 O5~V stars. The source of the ionization, then, either 
could be from a hot, luminous binary system or a compact cluster of a few massive stars.
The H~{\sc ii} region is not detected in a ground-based H$\alpha$ image from 2002 of the host 
galaxy\footnote{Available via NED.} obtained by the SINGS project \citep{Kennicutt+2003}.
The FWHM of the object is $\sim 0{\farcs}15$ at the WFPC2 scale, which, at the distance of NGC 
3938, is $\sim 12$--16\,pc.
For the sample of clusters in M74, those considered compact have effective radii $\lesssim 3$\,pc \citep{Ryon+2017}.
So, it is plausible that a massive progenitor system for SN 2017ein could be in the environment of a putative compact cluster and 
not be an actual member.
(Again, we note the 0.22 pixel, or 22 mas, offset between the SN and progenitor candidate positions in the {\sl HST\/} imaging,
which although within the measurement uncertainties, would represent an additional spatial offset of $\sim 1$--2\,pc.)
Such a progenitor system in a presumably actively star-forming environment along a spiral arm could still be massive, just not
necessarily from the same population and not necessarily of high mass. We are not in a position, based on the available data, to 
constrain such a system further.

An important test will be to return to the site years from now when the SN has faded, ideally either with {\sl HST\/} or the 
{\sl James Webb Space Telescope}, and determine whether the progenitor candidate still persists.
We note, however, that if the BPASS binary models are correct, a luminous secondary could still be evident at the SN location
(at the assumed distance range to the SN, a companion will not have perceptively drifted from its current position, as a 
result of a SN kick; the energy imparted on such a companion would also likely not be enough to modify its luminosity 
significantly), and the overall brightness of the object may have dimmed some, but not completely. From the BPASS models,
such a surviving companion would have apparent F814W $\approx 25.9$--24.6 and F555W$-$F814W 
$\approx 0.2$--0.4\,mag observed with WFC3/UVIS.

The inferred color and luminosity of the object may inevitably have been too blue and luminous to be physically realistic.  
The largest systematic uncertainty in the analysis is almost certainly in our estimate of the extinction to SN 2017ein.
The upper limit on the presence of the DIB $\lambda$5780
feature in the MMT spectrum, for instance, seems to indicate that $A_V$ could be $\lesssim 1.1$\,mag.
It is therefore important to better establish both the extinction and the distance to this SN. We have made the best effort given the
data we had available. It is possible that we have misestimated the reddening by using the intrinsic color curves for SN 2007gr 
and the reddening-free SN Ic color template from \citet{Stritzinger+2018}. Also, 
the host galaxy is likely beyond a practical distance for a tip-of-the-red-giant-branch (TRGB) measurement of its distance; 
however, the discovery and use of Cepheid variables via space-based observations is still possible.

Finally, we note in passing that the observations of SN 2017ein do not match well with the recent radiative-transfer models for
SESNe by \citet{Dessart+2016}. Although their relation between $M_R({\rm max})$ and
$M(^{56}{\rm Ni})$ predicts $\sim 0.06$--0.10 $M_{\odot}$, consistent with the results of our bolometric light-curve fitting
and $R$-band light-curve analysis, 
none of their model bolometric light curves provides a good match to that for SN 2017ein. 
Similarly, their model relation between the velocity at maximum light from the O~{\sc i}~$\lambda$7774 line 
and the SN expansion rate does not hold for this SN; at $R$-band maximum 
we measure a velocity in this line of $\sim 9800$\,km\,s$^{-1}$, which is well off the trend shown by 
\citet[][ their Figure 16]{Dessart+2016}.

\acknowledgments

We thank the referee for a careful review which improved this manuscript.
We also thank Ylva G\"otberg, Selma de Mink, and J. J. Eldridge for helpful discussions regarding SN Ic progenitor models, and
Monica Tosi regarding stellar clusters in nearby galaxies.
We appreciate comments on the paper provided by Jorick Vink, Scott Adams, Or Graur, and Kathryn Grasha.
Andrew Dolphin modified Dolphot to accept WFPC2 frames that have been flagged for cosmic-ray hits with
AstroDrizzle.
We are grateful to UC Berkeley undergraduate students
Sanyum Channa, Nick Choksi, Edward Falcon, Goni Halevi, 
Julia Hestenes, Ben Jeffers, and Samantha Stegman for helping
obtain some of the Lick Nickel data.
We kindly thank Thomas de Jaeger for assistance with some of the Kast observations.
Support for program GO-14645 was provided by the National Aeronautics
and Space Administration (NASA) through a grant from the
Space Telescope Science Institute (STScI), and this work  
is based in part on observations made with the NASA/ESA
{\it Hubble Space Telescope}, obtained from the Data Archive at 
STScI, which is operated by the Association of Universities for Research in Astronomy (AURA), Inc.,
under NASA contract NAS5-26555.
Partial support for N.S.'s supernova and transient research group at the University of Arizona was provided by NSF grant AST-1515559.
Support for A.V.F.'s supernova research group at U.C. Berkeley has            
been provided by U.S. NSF grant AST-1211916, the TABASGO Foundation,          
the Christopher R. Redlich Fund, and the Miller Institute for Basic           
Research in Science (U.C. Berkeley).       
This research has made use of the NASA/IPAC Extragalactic Database
(NED) which is operated by the Jet Propulsion Laboratory, California
Institute of Technology, under contract with NASA.
PyRAF is a product of the Space Telescope Science Institute, which is operated by AURA for NASA.

We thank the Lick Observatory staff for their expert assistance. KAIT and 
its ongoing operation were made possible by donations from Sun 
Microsystems, Inc., the Hewlett-Packard Company, AutoScope Corporation, 
Lick Observatory, the NSF, the University of California, the Sylvia \& Jim 
Katzman Foundation, and the TABASGO Foundation.  A major upgrade of the 
Kast spectrograph on the Shane 3~m telescope at Lick Observatory was made 
possible through generous gifts from the Heising-Simons Foundation as well 
as William and Marina Kast. 

Research at Lick Observatory is partially supported by a generous gift
from Google. We also greatly appreciate contributions from numerous
individuals, including Eliza Brown and Hal Candee, Kathy Burck and
Gilbert Montoya, David and Linda Cornfield, William and Phyllis
Draper, Luke Ellis and Laura Sawczuk, Alan and Gladys Hoefer, Roger
and Jody Lawler, DuBose and Nancy Montgomery, Rand Morimoto and Ana
Henderson, Jeanne and Sanford Robertson, Stanley and Miriam Schiffman,
Thomas and Alison Schneider, Mary-Lou Smulders and Nicholas Hodson,
Hans Spiller, Alan and Janet Stanford, the Hugh Stuart Center
Charitable Trust, Clark and Sharon Winslow, Weldon and Ruth Wood, and
many others.

\vspace{5mm}
\facilities{HST (WFPC2, WFC3), KAIT, Nickel, Shane (Kast Spectrograph), MMT (Blue Channel spectrograph)}

\vspace{5mm}
\software{
AstroDrizzle \citep[][  ~\url{http://drizzlepac.stsci.edu}]{Hack+2012},
Dolphot \citep{Dolphin2016}, DAOPHOT \citep{Stetson1987}, IRAF \citep{Tody1986,Tody1993}, 
PyRAF (\url{http://www.stsci.edu/institute/software_hardware/pyraf}),
extinction \citep[][ ~\url{https://github.com/kbarbary/extinction}]{Barbary2016},
galflat (\url{http://idlastro.gsfc.nasa.gov/ftp/pro/astro/gal_flat.pro}),\\
ishivvers/TheKastShiv (\url{https://github.com/ishivvers/TheKastShiv}),\\
pysynphot \citep[][ ~\url{https://github.com/spacetelescope/pysynphot}]{Lim+2015}
}

\allauthors

\listofchanges

\end{document}